\documentclass{article}
\usepackage{cite}
\usepackage{amsmath,amssymb,amsfonts}
\usepackage{algorithmic}
\usepackage{graphicx}
\usepackage{textcomp}
\def\BibTeX{{\rm B\kern-.05em{\sc i\kern-.025em b}\kern-.08em
    T\kern-.1667em\lower.7ex\hbox{E}\kern-.125emX}}

\usepackage{array} 
\usepackage{booktabs}
\usepackage{cleveref}
\usepackage{url}
\usepackage{ntheorem,paralist}
\usepackage{subcaption} 
\usepackage[verbose]{wrapfig} 
\usepackage{ntheorem}

\newenvironment{noinds_itemize}{\begin{list}{$\bullet$}
{\setlength{\rightmargin}{0em} \setlength{\leftmargin}{1.2em}
\setlength{\itemsep}{0em} \setlength{\topsep}{0em}
\setlength{\parsep}{0em}}}{\end{list}}

\newcommand\comment[1]{}
\newcommand\mypar[1]{{\bf #1.}}

{\theoremheaderfont{\itshape}
\theorembodyfont{\rmfamily}

}

\newcommand{\BL}{\mathrm{BL}} 
\newcommand{\LV}{\mathrm{LV}} 
\newcommand{\NR}{\mathrm{NR}} 
\newcommand{\Art}{5} 
\newcommand{\VN}{2} 
\newcommand{\LG}{1} 
\newcommand{\OT}{3} 
\newcommand{\KD}{4} 
\newcommand{\ES}{6} 
\newcommand{\LC}{7} 
\newcommand{\IV}{8} 
\newcommand{\DC}{9} 
\newcommand{\UR}{10} 
\newcommand{\MB}{11} 
\newcommand{\BE}{12} 
\newcommand{\LCBC}{13} 
\newcommand{\LCCD}{14} 
\newcommand{\BLPO}{15} 

\newcommand{\PT}{\mathrm{P}} 

\newcommand{\VLG}{\mathrm{V_{1}}}
\newcommand{\VAT}{\mathrm{V_{5}}}
\newcommand{\VVN}{\mathrm{V_{2}}}
\newcommand{\VOT}{\mathrm{V_{3}}}
\newcommand{\VKD}{\mathrm{V_{4}}}
\newcommand{\VESef}{\mathrm{V^{\prime}_{6}}}
\newcommand{\VLC}{\mathrm{V_{7}}}

\newcommand{\BF}{\mathrm{Q}} 
\newcommand{\QTL}{\mathrm{Q_{1}}}
\newcommand{\QOT}{\mathrm{Q_{3}}}
\newcommand{\QKD}{\mathrm{Q_{4}}}
\newcommand{\QLVVN}{\mathrm{Q_{2}}}
\newcommand{\QLVAT}{\mathrm{Q_{5}}}
\newcommand{\QESLC}{\mathrm{Q_{67}}}

\newcommand{\PTLG}{\mathrm{P_{1}}}
\newcommand{\PTOT}{\mathrm{P_{3}}}
\newcommand{\PTKD}{\mathrm{P_{4}}}
\newcommand{\PTES}{\mathrm{P_{6}}}
\newcommand{\PTESP}{\mathrm{P_{6p}}}
\newcommand{\PTLCP}{\mathrm{P_{7p}}}

\newcommand{\CL}{\mathrm{CL}} 
\newcommand{\CLcyp}{\mathrm{Q_{cyp}}} 

\newcommand{\dC}{\dot{C}} 

\newcommand{\fBL}{\mathrm{f_{b}}}
\newcommand{\fES}{\mathrm{f_{6}}}
\newcommand{\fLC}{\mathrm{f_{7}}}

\newcommand{\KmupES}{\mathrm{K_{a}}}
\newcommand{\KmpgpKD}{\mathrm{K_{b}}}
\newcommand{\KmpgpLC}{\mathrm{K_{b}}}
\newcommand{\KmmrpLC}{\mathrm{K_{c}}}

\newcommand{\VmaxupES}{\mathrm{J_{a}}}
\newcommand{\VmaxpgpKD}{\eta\mathrm{J_{b}}}
\newcommand{\VmaxpgpLC}{\mathrm{J_{b}}}
\newcommand{\VmaxmrpLC}{\mathrm{J_{c}}}
\newcommand{\eGFR}{\delta\mathrm{G_{fr}}}

\begin{document}
\title{Indirect Measurement of Hepatic Drug Clearance by Fitting Dynamical Models}
\author{Yoko Franchetti\footnote{Department of Pharmaceutical Sciences, Center for Clinical Pharmaceutical Sciences, University of Pittsburgh School of Pharmacy, Pittsburgh, PA 15261 USA (e-mail: yof8@pitt.edu)},
Thomas D. Nolin\footnote{Department of Pharmacy and Therapeutics, Center for Clinical Pharmaceutical Sciences, University of Pittsburgh School of Pharmacy, Pittsburgh, PA 15261 USA (e-mail: nolin@pitt.edu)
},
and Franz~Franchetti\footnote{Department of Electrical and Computer Engineering, Carnegie Mellon University, Pittsburgh, PA 15213 USA
}}
\maketitle

\begin{abstract}
We present an indirect signal processing-based measurement method for biological quantities in humans that cannot be directly measured. We develop the method by focusing on estimating hepatic enzyme and drug transporter activity through breath-biopsy samples clinically obtained via the erythromycin breath test (EBT): a small dose of radio-labeled drug is injected and the subsequent content of radio-labeled CO$_2$ is measured repeatedly in exhaled breath; the resulting time series is analyzed.  To model EBT we developed a 14-variable non-linear reduced order dynamical model that describes the behavior of the drug and its metabolites in the human body well enough to capture all biological phenomena of interest. Based on this system of coupled non-linear ordinary differential equations (ODEs) we treat the measurement problem as inverse problem: we estimate the ODE parameters of individual patients from the measured EBT time series. These estimates then provide a measurement of the liver activity of interest. The parameters are hard to estimate as the ODEs are stiff and the problem needs to be regularized to ensure stable convergence. We develop a formal operator framework to capture and treat the specific non-linearities present, and perform perturbation analysis to establish properties of the estimation procedure and its solution. Development of the method required 150,000 CPU hours at a supercomputing center, and a single production run takes CPU 24 hours. We introduce and analyze the method in the context of future precision dosing of drugs for vulnerable patients (e.g., oncology, nephrology, or pediatrics) to eventually ensure efficacy and avoid toxicity.
\end{abstract}

\section{Introduction}
This paper introduces a signal processing based indirect measurement method for
drug clearance activities of the liver in individuals, i.e., the elimination behavior of therapeutic drugs of interest. We provides the theoretical underpinning to estimate certain biological quantities that cannot be directly measured in live humans.
An additional complexity we address arises from multiple biological processes that may dependently overlap and thus cannot be easily separated. Our work is set in the context of future personalized medicine and precision medicine applications, and thus focusses on establishing the necessary methodological and mathematical foundations for attempting such future applications.

More specifically, we present an indirect measurement method for enzyme and drug transporter activity in the liver of individual patients using a single probe, called \textit{individualized physiologically based pharmacokinetic modeling of rate data (iPBPK-R)}.
We set up a reduced order physiologically based pharmacokinetic (PBPK) model of the human body, i.e., a system of 14 coupled non-linear ordinary differential equations (ODEs) where drug concentration in tissues of interest are modeled as state variables. The model is accurate enough to capture all biological effects of interest while keeping the number of model parameters low for stable parameter estimation. We introduce a graph representation and an operator formalism to concisely capture and analyze the particular non-linearities involved, and to perform perturbation analysis to assess the quality of the estimation procedure.

iPBPK-R solves an inverse problem to jointly estimate a set of ODE parameters that best fit a given measurement time series. These parameters provide indirect measurements for enzyme and transporter activity of interest. This is a hard optimization problem that needs to be carefully regularized via penalty terms and requires substantial computing resources, due the stiffness and non-linearity of the coupled ODEs and constraints imposed on the parameters by human biology. The method is implemented using R, and its development required 150,000 CPU hours on the \textit{Bridges} supercomputer at Pittsburgh Supercomputing Center (PSC) \cite{RN614}.

For the development of the iPBPK-R method we leveraged data obtained in previous clinical research \cite{RN728,RN19}, which was utilizing the erythromycin breath test (EBT): a small dose of the radio-labeled drug erythromycin is injected, and then the release rate of radio-labeled CO$_2$ in the patient's breath is measured at eleven time points over two hours. This resolves the initial transient, the saturated maximum, and the terminal slope of drug behavior in the human body. Further, release \textit{rate} of radioactive CO$_2$ carries higher information content than conventional concentration data. As an aside, the original EBT analysis procedure was simple and inconclusive, preventing clinical use of the EBT \cite{RN624}\cite{RN625}. We see potential future opportunities in utilizing a more advanced analysis method like iPBPK-R for not only EBT but other breath biomarker and probe development in this context since non-invasive breath biopsy research is emerging \cite{RN1000,RN1001}.

\mypar{Contribution}
This paper introduces and analyzes the mathematical and computational engineering framework behind iPBPK-R, our method to \textit{estimate} un-observable parameters related to drug clearance activity of the liver in \textit{individuals}. 
The iPBPK-R method applies a classical signal processing approach in a biological setting.
The purpose of this paper is to establish the mathematical and computational soundness of the method to enable future clinical research.

\begin{noinds_itemize}
\item We present both iPBPK-R's general mathematical framework and the particular
instance parameterized for the EBT as used in a previous clinical study \cite{RN728, RN19}.

\item iPBPK-R estimates biological quantities by solving a constrained optimization problem that fits a system of non-linear ODEs to clinical rate measurement data.

\item We introduce a graph representation of the coupled system of non-linear ODEs and an operator formalism to model and linearize its state-dependent adjacency matrix.

\item We utilize this formalism to provide a detailed mathematical analysis of the method, with focus on its constraints and
 the quality of its parameter estimates.

\end{noinds_itemize}

\noindent To provide context to the reader we summarize previously published applications of the method that estimate parameters related to drug clearance activity of the liver in individuals \cite{RN760} \cite{RN761}. Our method differs from the prevailing approach in the PBPK community since we do not predict drug behavior for a population, but \textit{estimate} parameters for \textit{individuals} given clinical measurement (breath) data. This paper is based on Chapter~2 of the dissertation thesis of Y Franchetti \cite{RN1040}.

\mypar{Synopsis} Sec.~\ref{sec:background} provides the necessary background. Sec.~\ref{sec:problem}--\ref{sec:procedure} develop the underlying mathematical and pharmacokinetic approach. Sec.~\ref{sec:experiments} shows clinical examples and applications while \cref{sec:discuss} discusses the method and related work, followed by a conclusion in \cref{sec:conclusions}.

\section{Background} 
\label{sec:background}

\begin{figure}[t!]
\vspace*{-3mm}
    \begin{subfigure}[t]{0.4\columnwidth}
        \centering
        \includegraphics[height=1in]{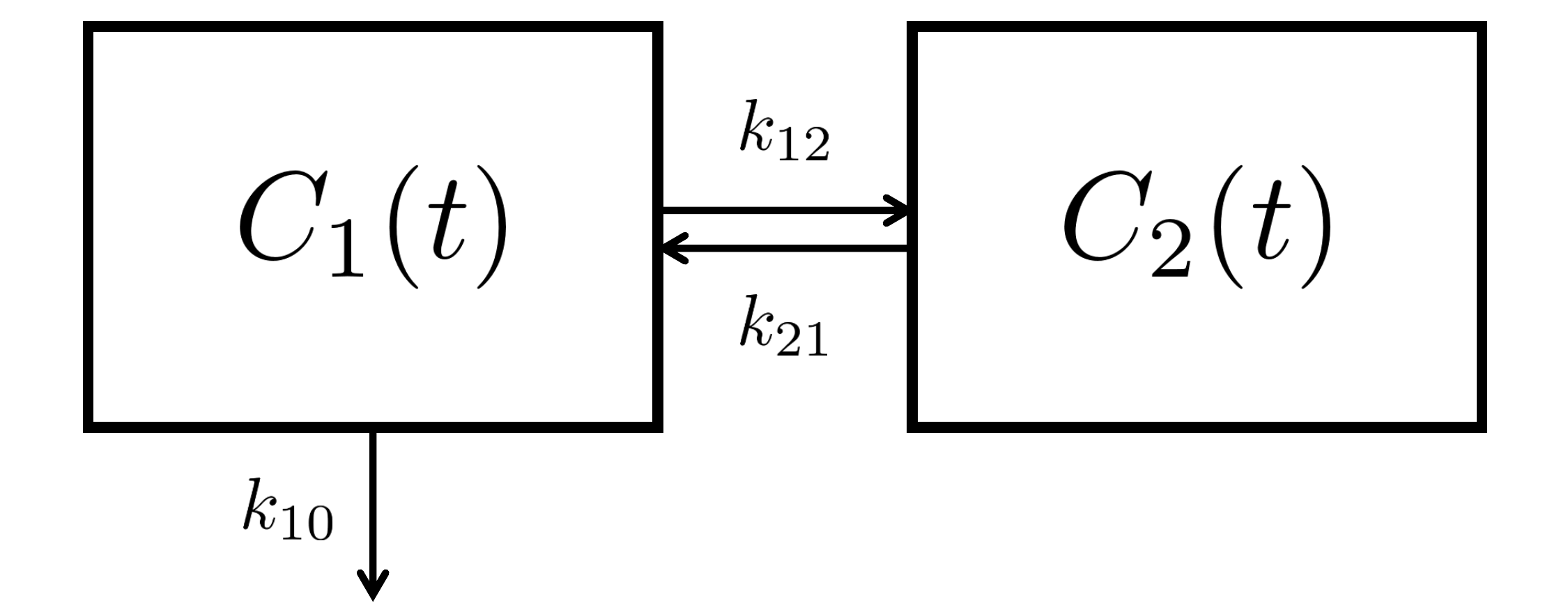}
        \caption{2-compartment model.\label{fig:twocomp}}
    \end{subfigure}
    \begin{subfigure}[t]{0.62\columnwidth}
        \centering
        \includegraphics[height=1.2in]{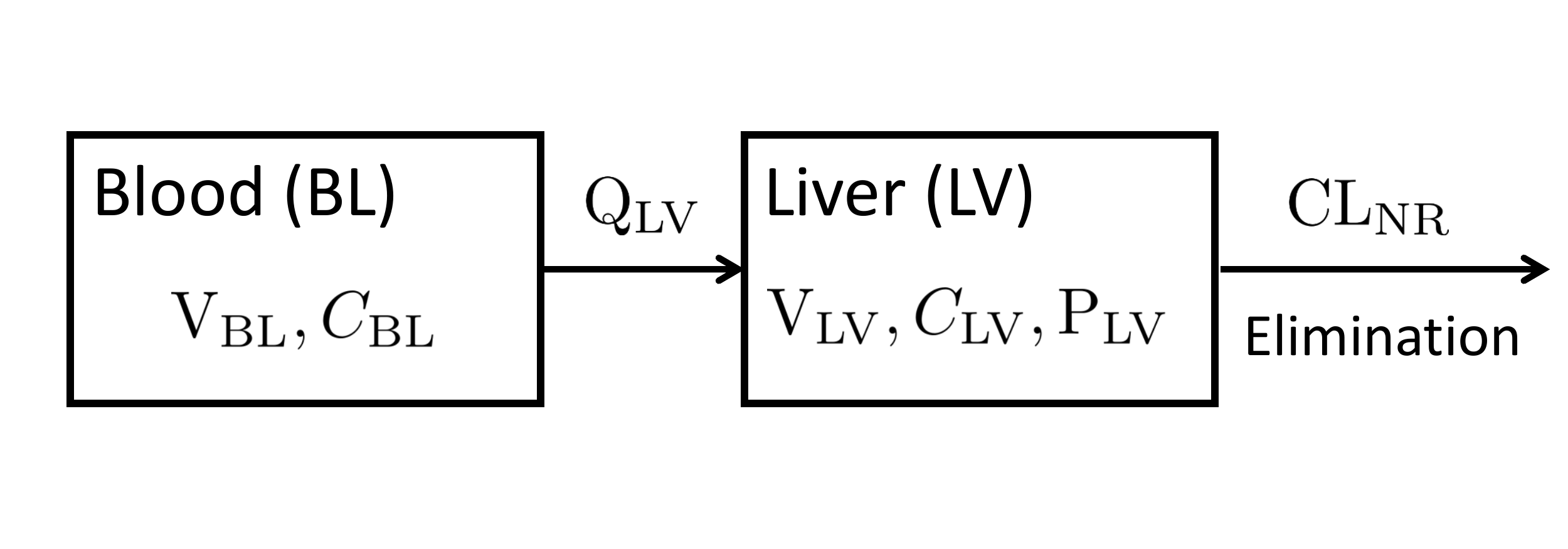}
        \caption{A simple PBPK model.\label{fig:simplepbpk}}
    \end{subfigure}
    \caption{Compartmental pharmacokinetic models.\vspace*{-3mm}}
\end{figure}

\subsection{Mathematical Background}

We briefly point to the mathematical concepts and methods used in this paper. The overall approach is following standard signal processing methodology~\cite{RN684}. We use graphs \cite{citeulike:395714} and non-linear ODEs \cite{RN892} to model the problem and non-linear operators \cite{RN893,RN764,RN765,RN896} to concisely describe the ODEs. We linearize these ODEs and use the theory of linear ODEs with constant coefficients to characterize the solutions via eigen decomposition of the system matrix \cite{RN897}. We use perturbation analysis \cite{RN898} to derive bounds and constraints.

We are estimating parameters of the ODEs to fit measurement data, as discussed in \cite{RN899,RN900,RN901,RN902,RN903}, and view the indirect measurement as inverse problem \cite{RN904}. The approach leads to non-linear optimization problems \cite{RN905,RN906} that need to be regularized \cite{RN872} and to be solved numerically \cite{RN907,RN908}. We implement the optimization in R \cite{RN909}.

\subsection{Pharmaceutical Science Background}
In the pharmaceutical industry PBPK modeling is typically used to predict drug concentration behavior over time. A system of ODEs is solved where each ODE is associated with a specific organ/tissue \textit{compartment} and describes the flows of drug concentrations for that compartment \cite{RN206}. In \textit{population PBPK}, physiologically meaningful parameters 
are incorporated in the ODEs. Then, they are used to predict drug concentrations of the respective compartments for a population \cite{RN674} \cite{RN209} \cite{RN416}. The prediction by PBPK modeling and simulation is informative for designing clinical trials and drug dosage. 

However, all predictive models raise concerns regarding performance in \textit{out-of-samples} settings 
where predictions 
 are made outside of the data range, and thus have statistical uncertainty \cite{RN751}. The physiological parameter 
inputs for these predictions are out-of-samples since 
equations to calculate the properties are typically developed based on multiple unrelated drug properties and thus extrapolate. The \textit{in-vitro in-vivo extrapolation (IVIVE)} described later is one such common calculation technique \cite{RN738} \cite{RN739} \cite{RN740}.

In contrast, our approach performs \textit{model fitting and parameter estimation}, not prediction. This change of viewpoint has a number of implications that we will discuss in \cref{sec:discuss}, chiefly that a smaller number of parameters is better to enable convergence to an unique solution, and that statistical considerations do not play a role. We now provide a quick overview of the relevant pharmaceutical science approaches.

\mypar{Pharmacokinetics (PK)}
PK is the quantitative study of drugs and their metabolites over time in the body \cite{RN632}. Compartmental models are  developed with some parameters related to absorption, distribution, metabolism, and excretion (ADME) of drugs to describe the time dependency of drug concentration in the body.
As example consider a two-compartment PK model for an intravenously dosed drug (see Fig.~\ref{fig:twocomp}), where the ODEs to describe the change in drug concentrations are given by 
$$
\dC_1 = -(k_{10}+k_{12})C_1+k_{21}C_2 \ \  \mathrm{and} \ \
\dC_2 = k_{12}C_1-k_{21}C_2.
$$
The concentration in Compartment $1$ is given by
$$
C_1(t) = P e^{-\alpha t} +  Q e^{-\beta t}
$$
where the constants
\begin{equation*}
P =  \frac{A_0 (\alpha - k_{21})}{\mathrm{V}_1 (\alpha - \beta)}
\quad \text{and}\quad
Q = \frac{A_0 (k_{21} - \beta)}{\mathrm{V}_1 (\alpha - \beta)}
\end{equation*}
depend on  $A_0$, $\alpha$, $\beta$, and $\mathrm{V}_1$ that capture administered drug mass, distribution rate constant, disposition rate constant, and volume of distribution of Compartment $1$, respectively.
\mypar{Physiologically-based PK (PBPK)} 
PBPK modeling was developed in the 1930s \cite{RN741} \cite{RN206}. The entire body is viewed as a system and each organ/tissue is viewed as a compartment. A system of ODEs for the entire body is solved to evaluate the drug concentrations over time for each compartment. Physiological parameters are included so that the impact of particular physiological behavior on drug kinetics in the body can
 be evaluated. PBPK models require three components: (1) system-specific properties, (2) drug properties, and (3) the 
structure of the system adapted to the research problem 
\cite{RN206}. 

See Fig.~\ref{fig:simplepbpk} for a simple PBPK model with blood (BL) and liver (LV) compartments and drug elimination from LV. The changes in drug concentration in BL and LV, 
\begin{align*}
\dC_{\BL}  &= \BF_{\LV}\big({\scriptstyle \frac{C_{\LV}}{\PT_{\LV}}} - C_{\BL}\big) \quad \mathrm{and}
\\
\dC_{\LV} &= \BF_{\LV}\big(C_{\BL}- {\scriptstyle \frac{C_{\LV}}{\PT_{\LV}}}\big) - \CL_{\NR}C_{\LV},
\end{align*}
form a system of ODEs where constants $\BF_{\LV}$, $\PT_{\LV}$ and $\CL_{\NR}$ capture the blood flow into LV, the partition coefficient of LV to BL, and the drug clearance from LV, respectively.

\mypar{IVIVE} 
IVIVE extrapolates the mean value of an \textit{in vivo} (clinically relevant) physiological parameter as a function of a corresponding \textit{in vitro} (experimentally/via bench work) obtained value  \cite{RN432} \cite{RN738}.
For example, metabolic clearance activity of CYP3A4 in human (\textit{in vivo} clearance) for a particular drug can be calculated using  \cite{RN716} \cite{RN698} as 
$$
\mathrm{CL_{H,int,3A4}} = \mathrm{CL_{int, rh3A4}}\, \mathrm{Abd_{3A4}}\, \mathrm{ISEF_{3A4}}\,\mathrm{MPPGL}\, \mathrm{LW}.
$$
Here, $\mathrm{CL_{int, rh3A4}}$ is the intrinsic clearance in recombinant CYP3A4 (\textit{in vitro} clearance)
, $\mathrm{Abd_{3A4}}$ is 
the unit amount of CYP3A4 enzyme
, $\mathrm{ISEF_{3A4}}$ is intersystem extrapolation factor for CYP3A4,
$\mathrm{MPPGL}$ is 
the unit amount of microsomal protein
, and $\mathrm{LW}$ is the human liver weight.

A value calculated via an IVIVE method is subject to substantial uncertainty. 
For example, experimentally determined ISEF for CYP3A4 clearance has a large $95$\% confidence interval (CI) 
\cite{RN739}
A calculated metabolic in vivo clearance 
may not have a strong correlation with experimentally determined in vitro metabolic clearance 
despite seemingly being linearly correlated in a log-log plot \cite{RN698}.
The relationship between IVIVE-predicted value and the 
observed \textit{in vivo} value is a unique feature of a particular drug substrate. IVIVE-based generalization of such a relationship to other drugs produces a bias in predictions as this is an \textit{out-of-samples} extrapolation.

Scaling factors are commonly derived from regression in a log-log plot, which does not predict in vivo metabolic clearance consistently. Therefore, it is not possible to estimate \textit{a particular subject's} in vivo parameters using an IVIVE-derived scaling factor given the low predictive power and great uncertainty of this scaling factor.
In contrast, we estimate an \textit{adjustment factor} via PBPK modeling that measures an individual's deviation from the baseline IVIVE activity to \textit{estimate} the individual's activity of a biological process.

\subsection{Erythromycin Breath Test (EBT)}\label{subsec:ebt}
\begin{figure}[t!]
\vspace*{-0mm}
\begin{subfigure}[t]{0.5\textwidth}
        \includegraphics[height=2in]{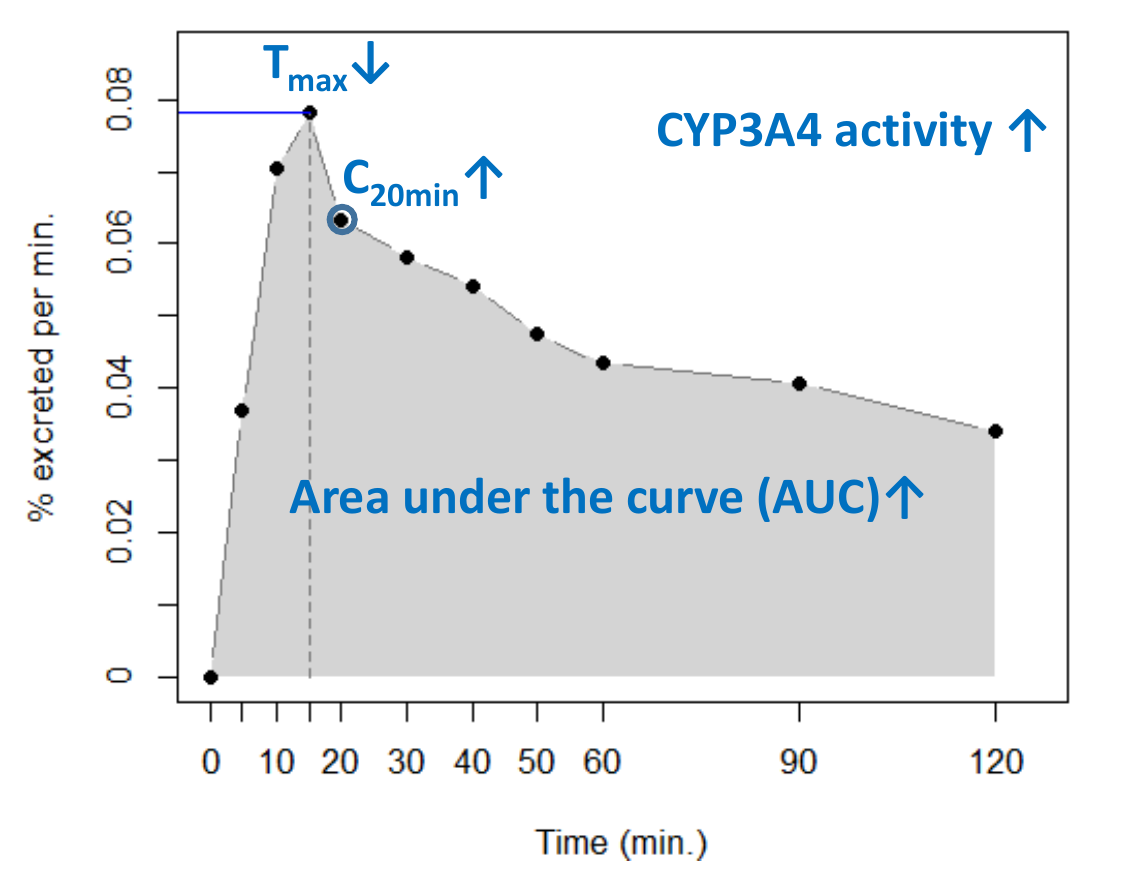}
        \caption{Interpretations of EBT data.\label{fig:oldebt}}

    \end{subfigure}%
    \begin{subfigure}[t]{0.5\textwidth}
        \hspace*{-5mm}
        \includegraphics[height=1.5in]{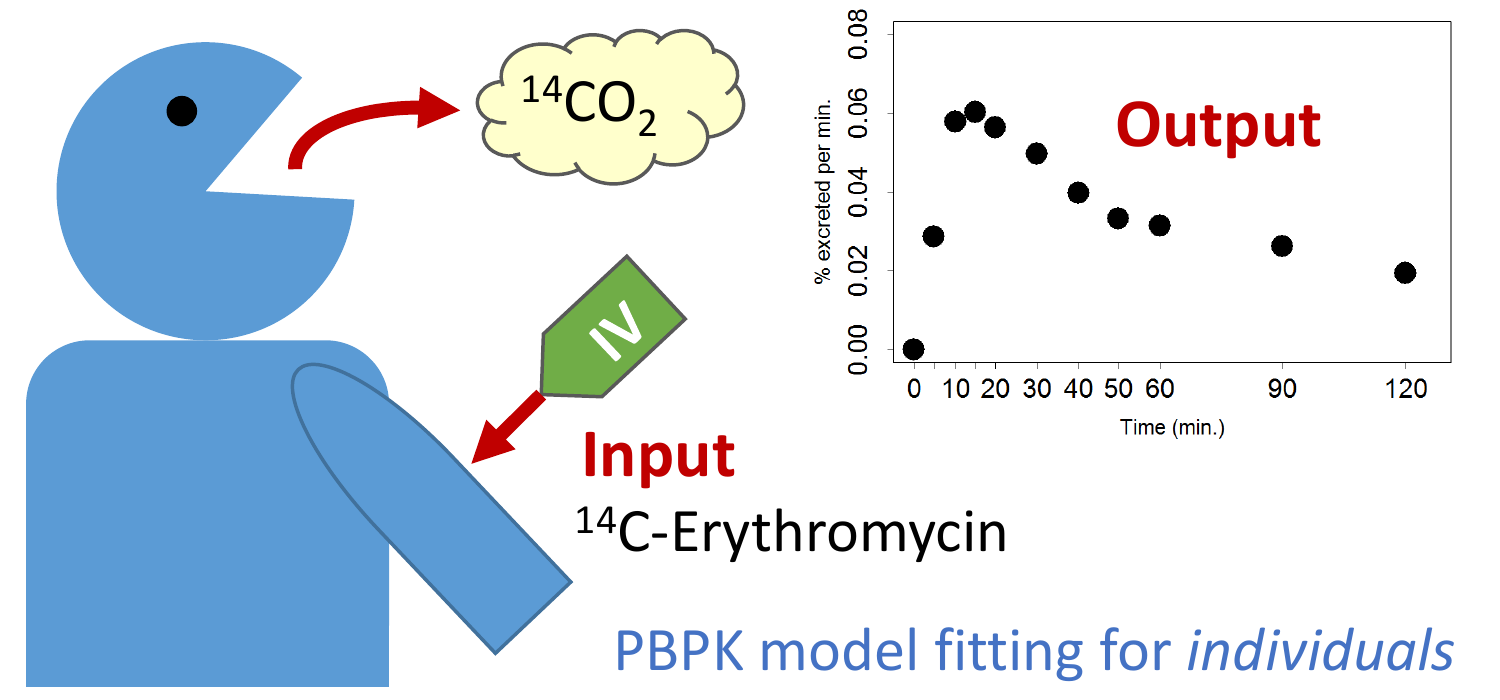}
        \caption{EBT procedure.\label{fig:ebt}}
    \end{subfigure}%
    \caption{EBT and its classical interpretations.\vspace*{-0mm}}
\end{figure}

EBT was originally developed to measure 
CYP3A4 activity that metabolizes the radio-labeled drug $^{14}$C-erythromycin in the liver \cite{RN618}. EBT is based on the premise that as the drug undergoes the CYP3A4-mediated metabolic pathway, 
radio-labeled carbon dioxide ($^{14}$CO$_2$) is released as a final by-product. 
A subject receives a noninvasive, single intravenous (IV) bolus low dose of $^{14}$C-erythromycin. 
Then, breath samples are collected at 11 time points within two hours of IV dosing, including the baseline time point (see Fig.~\ref{fig:ebt}) \cite{RN19} \cite{RN728}. The sampling time points are unequally spaced
. $^{14}$CO$_2$ production rates in the breath samples are calculated (as a percentage of dose exhaled per minute) at each time point from the exhaled volume and $^{14}$CO$_2$ content.

In the original EBT procedure breath samples are collected
before and after a certain change in the subject's state. The resulting time series is used to detect \textit{binary evidence (yes/no) of increase in CYP enzyme activity} as follows (see Fig.~\ref{fig:oldebt}): the CYP3A4 enzyme activity in the liver \textit{increased} \textit{if} (1) the time to peak of the breath rate 
shortened, \textit{or} (2) the area under the curve increased, \textit{or} (3) the measured $^{14}$CO$_2$ production rate at $20$ minutes increased post the state change.

However, other physiological activities 
also seemed to affect elimination of the drug from the liver (e.g., activities in drug transporters) \cite{RN60} \cite{RN440} \cite{RN406} \cite{RN217} \cite{RN620}, and the original EBT criteria were inconclusively predicting CYP activity changes
\cite{RN624} \cite{RN625}. iPBPK-R applied to EBT is revisiting the EBT analysis by taking advantage of $^{14}$CO$_2$ production rate data.
Our approach suggests that more detailed modeling of EBT data can provide more granular information regarding CYP3A4 activity and other physiological factors.

\section{Mathematical Framework of iPBPK-R}
\label{sec:problem}
\subsection{Overview}
The key idea of iPBPK-R is to  develop an {\em indirect measurement method} to simultaneously measure values in a particular patient for physiological parameters that cannot be measured directly.
From an applied statistics perspective we are estimating parameters in a \textit{within sampling} setting as opposed to \textit{out-of-samples} setting: given clinically observed breath rate data from a single probe drug (i.e., EBT), PBPK model fitting (but not prediction) will be used to inversely solve for physiologically meaningful parameters explaining the response.

iPBPK-R is based on a reduced order PBPK model, which is parameterized by our physiological parameters of interest.
Using measured breath rate data allows us to resolve the early transient (the impulse response) in the drug behavior at both rate-limiting and non-rate-limiting steps of multiple elimination pathways.
IVIVE parameters are utilized as initial guesses that are adjusted for individual subjects based on their measurement data.
Estimation is done via nested optimization or nested co-optimization and utilizes a specialized objective function (loss function).
iPBPK-R enables us to estimate per-person physiological parameters which are otherwise difficult to capture, by
utilizing observed concentration \textit{change} data (production \textit{rate} data), which at the same sampling rate and accuracy has higher information contents compared to drug \textit{concentration} data.
We develop the mathematical framework as follows.
In \cref{subsec:reduced} we introduce a graph representation of the non-linear system of ODEs. Then
in \cref{subsec:matrixODEs} we introduce an operator formalism based on \cite{RN765} \cite{RN764} to capture the non-linearities in the model concisely, and in \cref{subsec:Ltwonorm} we formalize measurements and parameter estimation as optimization problem. This enables us to establish convergence properties in \cref{subsec:accuracy} via perturbation analysis.

\subsection{Reduced Model}
\label{subsec:reduced}
\begin{figure}[h!]
\centering
\includegraphics[width=80mm]{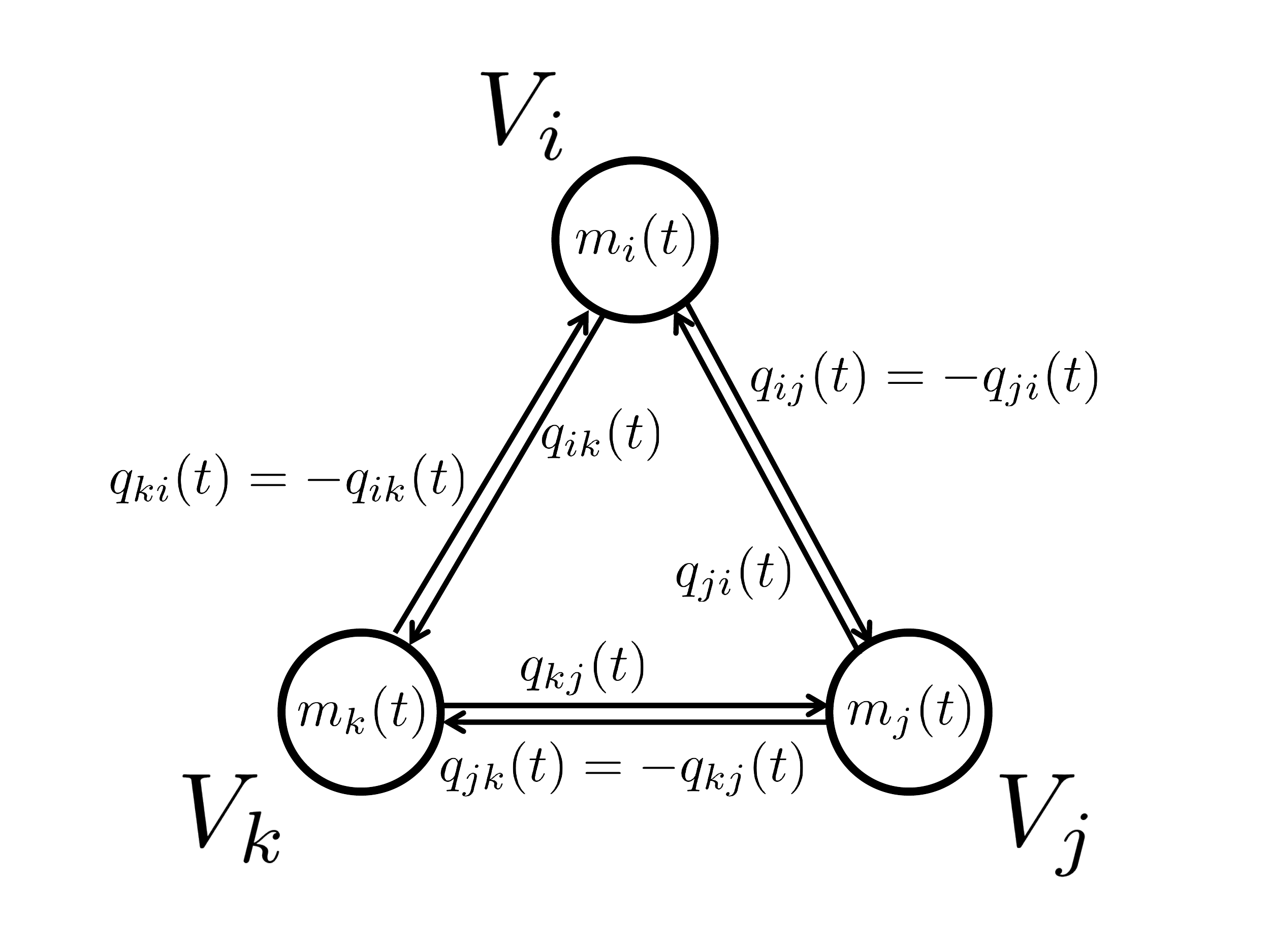}
\caption{Example graph for a three compartment model.\label{fig:model}}
\vspace{-5mm}
\end{figure}

In this subsection we describe the general shape of our reduced model. This model is underlying the modeling approach and is instantiated for EBT in
\cref{sec:mtd}. Fig.~\ref{fig:model} shows a simple three compartment model and all associated quantities used in the discussion below.

\mypar{Compartments}
The reduced model underlying iPBPK-R is given by a system of non-linear ordinary differential equations (ODEs) for time $t \geq 0$,
described as a weighted directed graph (see Fig.~\ref{fig:model})
\begin{equation}\label{eq:model}
\mathrm{G} = (\mathrm{V}, \mathrm{E}, \mathrm{W})
\quad\text{with}\quad
\mathrm{E} \subseteq \mathrm{V} \times \mathrm{V}.
\end{equation}
The set of $n$ vertices
$\mathrm{V} =\{V_1,\dots,V_n\}$
abstracts the compartments. Each compartment $V_i$ has a state variable
\begin{equation}\nonumber
m_i(t),\ t\geq 0
\quad \text{with}\quad
m_i(0) =M_i
\end{equation}
that is a function of time and expresses the amount of drug mass in compartment $V_i$ at time $t$.

\mypar{Mass flows}
Edges $e_{ij}=(V_i,V_j)$ and $e_{ji}=(V_j,V_i)$ in $E$
together represent the channel between compartments $V_i$ and $V_j$, viewed from the respective compartment. The weight $w_{ij}\in\mathrm{W}$ on edge $e_{ij}$ is given by the mass flow function
\begin{equation}\nonumber
q_{ij}(t)
= \dot{m_i}(t)
 \quad\text{for}\quad e_{ij} \in \mathrm{E},
\end{equation}
which expresses the flow of drug mass from compartment $V_i$ into $V_j$ over the edge $e_{ij}$ as function of time $t$.
We define
$
q_{k\ell}(t) \equiv 0 \quad\text{if}\quad e_{k\ell} \notin \mathrm{E},
$
i.e., for edges that do not exist.
$q_{ij}(t)$ and $q_{ji}(t) = - q_{ij}(t)$ denote the same physical flow represented from the perspective of compartment $V_i$ and $V_j$, respectively,  as flow source.
Thus the graph $\mathrm{G}$ has a $n \times n$  real-valued time-dependent anti-symmetric adjacency matrix
\begin{equation}\label{eq:odemat}
A(t) = [q_{ij}(t)]_{1\leq i,j \leq n}
\end{equation}
that concisely captures the system of non-linear ODEs of the reduced model.

\mypar{State equations and invariants}
Note that in our model some compartments $V_i$ are {\em sources} for which all flows $q_{ij}\geq0$ are non-negative (thus, they are out-flows),
and some compartments $V_i$ are {\em sinks} for which all flows $q_{ji}\geq0$ are non-negative (thus, they are in-flows).
The non-linearity of the model arises through non-linearities of certain flow terms $q_{ij}(t)$ that are induced through a Michaelis-Menten style saturation (see \cref{subsec:matrixODEs}). 

For each compartment $V_i$ the change of drug mass contents is the sum of all flows $q_{ji}(t)$ from all other compartments $V_j$ into $V_i$, and there are no self-flows. Thus,
\begin{equation} \label{eq:sumq}
\dot{m_i}(t) = \sum_{j=1}^{n} q_{ji}(t)
\ \ \text{and}\ \
\sum_{i=1}^{n} m_i(t) =
\sum_{i=1}^{n} M_i = M_0.
\end{equation}
Since mass can only enter or leave a compartment $V_i$ through a mass flow $q_{ij}$, the total mass in the system is constant $M_0$.

\mypar{Concentration based model}
In Clinical Pharmacology it is common to express models via drug concentration and its time-dependent change
$$
C_i(t)=\frac{m_i(t)}{V_i}
\quad\text{and}\quad
\dot{C}_i(t)=\frac{\dot{m}_i(t)}{V_i},
$$
respectively (here $V_i$ denotes the volume of the $i$th compartment), while mass-based equations allow for clearer mathematical treatment of the ODEs.
Thus, in pharmacological discussions we will use $C(t)$ and $\dot{C}(t)$ while mathematical discussions will use $m(t)$ and $\dot{m}(t)$, respectively.

\subsection{Reduced Model as Non-linear ODE System}
\label{subsec:matrixODEs}
We now define a framework that allows us to describe the system of non-linear ODEs \eqref{eq:sumq} via the time-dependent adjacency matrix \eqref{eq:odemat} similar to  how a system of linear first order ODEs with constant coefficients is described through its constant system matrix.

\mypar{Definitions and conventions} 
When needed to disambiguate vectors and scalars, vectors $x\in\mathbb{R}^n$ are annotated as $\vec{x}$. Entries of the vector $x$ are denoted as $x_i$,  and the vector $x$ can be written as $x = (x_i)_{1 \leq i \leq n}$.  Real $m \times n$ matrices are denoted as $\mathbf{A} = [a_{ij}]_{1 \leq i \leq m,\, 1 \leq j \leq n}$. 

The \textit{Michaelis-Menten} function $\mathrm{MM}_{\mathrm{a,b,c}}$ is a homogeneous hyperbolic function that is parameterized by three constants $a$,$b$, and $c$,
\begin{equation}\label{eq:mmf}
            \mathrm{MM}_{\mathrm{a,b,c}}(x) = \frac{ax}{b+cx}.
\end{equation}
It is the main non-linearity used in the reduced model, and it is a monotonically increasing differentiable function.
The parameters capture the initial slope and the asymptotic value of the function.
We denote a function $f(.)$ that depends on a parameter vector $p$ as $f^p(.)$.
In our case the parameters $p_i$ that constitute the vector $p$ are to be optimized during estimation.

\mypar{Notation for non-linear operators}
We unify the notation for linear and non-linear operators
$
\mathbf{A}(.)
$
that map vectors to vectors.
This will allow us to analyze and linearize the non-linear system of ODEs \eqref{eq:sumq}, and to obtain a good characterization of the original non-linear system with respect to parameter estimation.
A matrix $\mathbf{A} \in \mathbb{R}^{m \times n}$ defines a linear operator
$$
\mathbf{A}(.) : \mathbb{R}^n \to \mathbb{R}^m;\, x \mapsto \mathbf{A}(x) = \mathbf{A}x,
$$
and we use the notation $\mathbf{A}(.)$ for a matrix $\mathbf{A}$ to denote its interpretation as operator.

We next generalize the matrix-vector product to matrices where the entries are scalar functions.
Assume that entries at location $(i,j)$ of a $m \times n$ matrix $\mathbf{B}$ are scalar functions
\begin{equation}
b_{ij}(.) : \mathbb{R} \rightarrow \mathbb{R};\, x \mapsto b_{ij}(x).
\end{equation}
We use $\mathbf{B}$ to define the \textit{non-linear operator} $\mathbf{B}(.)$ as
\begin{equation}\label{eq:operator}
\mathbf{B}(.) : \mathbb{R}^n \to \mathbb{R}^m;\,
\mathbf{B}(x) = y
\quad\text{with}\quad
y_i = \sum_{j=1}^{n} b_{ij}(x_j). 
\end{equation}
The definition of $\mathbf{B}(.)$ generalizes the standard matrix-vector product.
Finally, we define operator addition $\mathbf{U}(.)+\mathbf{V}(.)$ as
\begin{equation}\label{eq:addop}
\big(\mathbf{U}(.)+\mathbf{V}(.)\big)(x) = \mathbf{U}(x) + \mathbf{V}(x),
\end{equation}
compatible with the usual addition of matrices.

\mypar{Reduced model ODE as non-linear operator} We now describe the non-linear system of ODEs concisely as a sum of operators.
Our state vector is the vector of mass in all compartments of the $n$-compartment model, $\vec{m} = (m_i)_{1 \leq i \leq n}$.
The forcing function $z(t)$ is zero everywhere except for the $k$th component $z_k(t) = \tau$ in the interval $[0,t_0]$
to model intravenous (IV) injection into the vein compartment.

The linear part of the system of ODEs is given by a real $n \times n$ matrix $\mathbf{X}$. All non-linearities are collected in the $n \times n$ matrix of scalar functions $\mathbf{Y}=\big[y_{ij}(.)\big]_{i,j}$ where all entries $y_{ij}(.)$ are either the \textit{zero function} $o(x)\equiv0$, a single Michelis-Menten function \eqref{eq:mmf}, or a sum of two Michelis-Menten functions \eqref{eq:mmf}.
The matrix of functions $\mathbf{Y}$ is used to describe the operator $\mathbf{Y}(.)$ as defined in \eqref{eq:operator}.
With these definitions, the full non-linear system of ODEs is given by
\begin{equation}\label{eq:odeconc}
\dot{m} = \mathbf{M}(m) + z(t)
\quad\text{with}\quad
\mathbf{M}(.) =
\mathbf{X}(.) + \mathbf{Y}(.).
\end{equation}
$\mathbf{M}(.)$ is the state-dependent adjacency matrix \eqref{eq:odemat}. Analysis of the solution of \eqref{eq:odeconc} and convergence and uniqueness of parameter estimation via optimization is reduced to analyzing $\mathbf{M}$. In addition, the non-linear behavior of $\mathbf{M}$ is limited to Michaelis-Menten saturation. This enables us to analyze solution properties of \eqref{eq:odeconc} via perturbation analysis of the entries of $\mathbf{M}$, and by analyzing the linearized bounds of $\mathbf{M}$.

\subsection{Measurements and Parameter Estimation}\label{subsec:Ltwonorm}

\mypar{Measurement} Next we model the pathway of the drug by-product $^{14}$CO$_2$ from the liver to the breath and its ultimate concentration measurement in the breath.
This is described as a measurement operator  $\mathcal{F}(.)$ that maps the solution vector $\vec{C}(t)$ of \eqref{eq:odeconc} (in drug concentration form) to a scalar function $B(t)$. Under the assumption of instant metabolism of the drug in the liver and instant transfer of $^{14}$CO$_2$ from liver to lung, the operator $\mathcal{F}(.)$ just returns the first derivative
of the liver compartment (denoted as $j$th compartment), i.e.,
\begin{equation}\label{eq:breath}
B(t) = \mathcal{F}(\vec{C}(t)) = \dot{C_j}(t).
\end{equation}
More realistic variants of $\mathcal{F}(.)$ can be defined to model the transfer of $^{14}$CO$_2$ from the liver to the lung more accurately, and we provide further details in \cref{sec:procedure}.

\mypar{Comparing model and measurement}
Next we define a distance function from $B(t)$ as defined in \eqref{eq:breath} to clinical breath rate measurements. A measurement consist of $T$ samples $m_\ell=(t_\ell, w_\ell)$ measuring $^{14}$CO$_2$ production rate taken at time points $t_\ell$, arranged as a vector of 2D points, $\vec{S} = \big((t_\ell, w_\ell)\big)_{\ell}$ for $\ell=1,\dots,T$.
We define a \textit{distance function} 
$d(.,.)$ 
to denote the distance (or disagreement)
between the simulated function and the sampled data.
A straight-forward example for a distance function $d(.,.)$ is the $\mathrm{L}_2$ norm of the breath function $B(t)$ evaluated at the sample time points $t_\ell$ minus the measured data at the same time points,
\begin{equation}\label{eq:dltwonorm}
d_2(B(.), \vec{S}) = \| (B(t_\ell))_{\ell} - (w_\ell)_{\ell}  \|_2
,\quad\ell=1,\dots,T,
\end{equation}
which is provided to aid the discussion regarding parameter estimation. We present the more complex definition and detailed discussion of the actually used distance functions in \cref{sec:procedure}. 

\mypar{Parameter estimation via optimization}
The estimation of biological parameters is cast as an optimization problem that fits a dynamical model (system of ODEs) to measurement data. The biological parameters of interest are derived from the estimated ODE parameters that produce the best fit.
We aim to find a vector $\vec{r}$ that parameterizes the system of ODEs \eqref{eq:odeconc} (in concentration form) so that the
distance $d(.,.)$ from a given measurement vector $\vec{S}$ is minimized. Further, constant parameters are captured by a parameter vector $\vec{u}$. Therefore the flow terms $q_{ij}(t)$ in \eqref{eq:sumq} and the solution of this system of ODEs $\vec{C}(t)$
are parameterized by $\vec{r}$ and $\vec{u}$ and need to be written as
\begin{equation*}
q_{ij}^{\vec{r},\vec{u}}(t)
\quad\text{and}\quad
\vec{C}^{\vec{r},\vec{u}}(t),
\end{equation*}
respectively.
The parameter estimation is set up as the optimization problem
\begin{equation}\label{eq:opt}
\vec{r}
= \underset{\vec{r}'}{\arg\min}\,\varPsi^{\vec{u}}(\vec{r}')
\quad\text{for configuration }\vec{u}
\end{equation}
for a specialized objective function
\begin{equation}\label{eq:objf}
\varPsi^{\vec{u}}(\vec{r})= d(\mathcal{F}(\vec{C}^{\vec{r},\vec{u}}(.)),\vec{S})+\pi(\vec{r}, \vec{u})
\end{equation}
with appropriately chosen distance $d(.,.)$ and penalty term $\pi(\vec{r}, \vec{u})$.
To ensure the that the estimation returns biologically plausible values and adjustment factors are as close to 1 as possible, the penalty function $\pi(.,.)$ needs to be chosen carefully.
The penalty term is defined as
\begin{equation}\label{eq:pi}
\pi(\vec{r}, \vec{u}) =
 u_\mathrm{b}\nu(\vec{r}) + u_\mathrm{c}\rho(\vec{r}) + u_\mathrm{d}\epsilon(\vec{r}) + u_\mathrm{x}\psi(\vec{r}).
\end{equation}
The terms $\nu(\vec{r})$, $\rho(\vec{r})$, $\epsilon(\vec{r})$, and $\psi(\vec{r})$ are constraint or penalty terms that depend on the parameter vector $\vec{r}$, and the constants 
$u_\mathrm{b}$, $u_\mathrm{c}$, $u_\mathrm{d}$, and $u_\mathrm{x}$ are weighting factors that are collected in the configuration vector $\vec{u}$.
The exact forms of the penalty terms will be discussed in \cref{sec:procedure}.

\subsection{Analysis of iPBPK-R Estimation}\label{subsec:accuracy}

In this section we establish that parameter estimation (as it is set up in \cref{subsec:Ltwonorm}) is converging as long as good starting values (in our case IVIVE estimates) are known and the number of estimated parameters is properly bounded.

\mypar{Linearization of the system of ODEs}
The nonlinear ODE system \cref{eq:odeconc} can be bounded as
\begin{equation}\nonumber 
             (\mathbf{X} + \mathbf{Y}^{-})\vec{m}(t) + \vec{z}(t)
             \preceq
\frac{d\vec{m}(t)}{dt}
\preceq
             (\mathbf{X} + \mathbf{Y}^{+})\vec{m}(t) + \vec{z}(t)
\end{equation}
\noindent ($\preceq$ denotes {\em element-wise} comparison) where
\begin{align*}
            \mathbf{Y}^{-} &= \big[y_{ij}^-\big]_{i,j=1,\dots,n} \quad \mathrm{with} \quad y_{ij}^- = \underset{t \geq 0}{\min}\, \dot{y}_{ij}\big(m_j(t)\big)
            \quad \text{and}
\\
            \mathbf{Y}^{+} &= \big[y_{ij}^-\big]_{i,j=1,\dots,n} \quad \mathrm{with} \quad y_{ij}^+ = \underset{t \geq 0}{\max}\, \dot{y}_{ij}\big(m_j(t)\big).
\end{align*}
\noindent $\mathbf{Y}^{-}$ and $\mathbf{Y}^{+}$ are matrices representing elementwise upper and lower bounds on the matrix elements of $\mathbf{Y}$. Therefore,
\begin{equation}\label{eq:boxY}
            y^{-}_{ij}m_j(t) \, \le \, y_{ij}\big(m_j(t)\big) \, \le \, y^{+}_{ij}m_j(t)
\end{equation}
\noindent holds for all matrix elements of $\mathbf{Y}$ for all $t\geq 0$.
For iPBPK-R to produce stable parameter estimates, the distance between the boundaries $|y^{+}_{ij} - y^{-}_{ij}|$ needs to be small enough.
Then arguments for the linearized upper and lower bounds hold for the non-linear system of ODEs.
The exact bounds are problem dependent and can be checked post-hoc.

\mypar{Solution of linearized ODE} The solution $\vec{x}(t)$ of a homogeneous system of $n$ linear ODEs with a {\em simple} real system matrix $\mathbf{A}$ (which has $n$ linearly independent eigenvectors),
\begin{equation}\label{eq:simplelinearODE}
            \frac{d\vec{x}(t)}{dt} = \mathbf{A}\vec{x}(t), \,\quad \vec{x}(t) \in \mathbb{R}^{n},\quad \mathbf{A}\in\mathbb{R}^{n \times n},\quad t \geq 0
\end{equation}
is given by
\begin{equation}\label{eq:linearODEsolvector}
            \vec{x}(t) = \sum_{i=1}^{n}\beta_i\vec{c}_i e^{\theta_i t} \quad \text{with}\quad \vec{x}(t)=(x_j(t))_{j=1,\dots,n}.
\end{equation}
We denote eigenvalues and eigenvectors of the system matrix $\mathbf{A}$ as $\theta_i$ and $\vec{c}_i$, respectively, and the only eigenvalue with multiplicity greater than 1 is $\theta_0=0$.
In our application $\mathbf{A}^{-} = (\mathbf{X} + \mathbf{Y}^{-})$ and $\mathbf{A}^{+} = (\mathbf{X} + \mathbf{Y}^{+})$, and we ignore the forcing function $\vec{z}(t)$ since $\vec{z}(t) \equiv0$ for $t > t_0$.
\noindent The $j$th element of $\vec{x}(t)$ is the scalar solution for the $j$th compartment,
\begin{equation}\nonumber 
            x_j(t) 
             = \sum_{i=1}^{n} \beta_ic_{j,i} e^{\theta_i t} \quad \mathrm{where} \quad
             \vec{c}_i = (c_{j,i})_{j=1,\dots,n}.
\end{equation}
The function $x_j(t)$ is the projection of $\vec{x}(t)$ into the $j$th dimension (i.e., $j$th compartment).

\mypar{Construction of ODE solution from projection}
We now derive the conditions under which \textit{the entire solution} $\vec{x}(t)$ of \cref{eq:simplelinearODE} can be reconstructed from the $j$th dimension.
A measurement consists of $T$ samples $(t_{\ell},\, x_j(t_{\ell}))_{{\ell}}$ of the function $x_j(t)$.
Under our assumptions the (normalized) eigenvectors $\vec{c}_i$ are
functions of the eigenvalues $\theta_i$.
This allows us to set up a non-linear system of $2n$ equations in $\beta_i$ and $\theta_i$,
\begin{equation}\label{eq:linearODEsolxjobs}
            x_j(t_{\ell})
             = \sum_{i=1}^{n} \beta_ic_{j,i}(\theta_i) e^{\theta_i t_{\ell}}
             \quad\text{for}\quad \ell = 1,\dots, T.
\end{equation}
Thus, at least $2n$ observed samples are necessary to estimate the $2n$ unknown values $\beta_{i}$ and $\theta_i$. More samples lead to better estimates via solving a non-linear least squares problem.
Higher multiplicity of $\theta_0=0$ does not pose a problem as the nullspace of $\mathbf{A}$ captures the source and sink of the dynamical system for which we do not need unique solutions.

\mypar{Perturbation of ODE for the linearized system}
We define a small perturbation $\alpha = (1+\varepsilon) \approx 1$ for a matrix entry $a_{ij}$, i.e., replace $a_{ij}$ with $\alpha a_{ij}$ and view the resulting matrix $\mathbf{\tilde{A}}$ as a function of $\alpha$, i.e., $\mathbf{\tilde{A}}(\alpha)$.
For small enough $\alpha$ the eigenvalues of $\mathbf{\tilde{A}}$ depend continuously and differentiably on $\alpha$.
This can be shown by using Laplace's formula for the determinant, Cramer's solution formula for linear systems, and the fundamental theorem of algebra.
Thus, for a small enough $\alpha$ the partial derivatives
\begin{equation*}
\frac{\partial \theta_i}{\partial \alpha}
\quad \mathrm{and} \quad
\frac{\partial \beta_i}{\partial \alpha}
\end{equation*}
\noindent exist at least locally
under weak assumptions that are met in practice by iPBPK-R.
The above discussion generalizes to multiple perturbations $\alpha_i$.

\mypar{Parameter estimation for the linearized system}
The set of $T$ non-linear equations \eqref{eq:linearODEsolxjobs} can be recast as non-linear least squares problem for
$\vec{\alpha} = (\alpha_1,\dots,\alpha_M)$,
\begin{equation}\nonumber 
\vec{\alpha} =
\underset{\tilde{\alpha}\in\mathbb{R}^{M}}{\arg\min}\, \sum_{\ell = 1}^T
 \Big(           x_j(t_{\ell})
-
\sum_{i=1}^{n} \beta_i(\tilde{\alpha})c_{j,i}(\tilde{\alpha}) e^{\theta_i(\tilde{\alpha}) t_{\ell}}\Big)^2
 \end{equation}
that has at least one solution that can be found for initial values close to the optimum.
Thus, the adjusted system matrix $\tilde{\mathbf{A}}(\vec{\alpha})$ needs to be close to the matrix $\mathbf{A}$, which is derived from IVIVE parameters. The local minimum $\vec{\alpha}$ closest to $\mathbf{1}_M=(1,\dots,1)$ is the most \textit{biologically plausible} estimate.
Since $\theta_i$ are the roots of the characteristic polynomial of \cref{eq:simplelinearODE} of degree $n$, at maximum $n$ independent perturbations $\alpha_i$ can be estimated, and at least $2n$ data points $x_j(t_\ell)$ are needed.

\mypar{Parameter estimation for nonlinearity} 
Recall that all non-linearities are of Michaelis-Menten form and have two true parameters (initial slope at 0 and maximum value at infinity).
The nonlinear Michaelis-Menten saturation suppresses the peaks of the exponential terms in the solution $\vec{x}(t)$ and thus the shapes of the solution-time curves deviate from those of the corresponding pure exponential terms, in particular for high values of $\|\vec{x}(t)\|$. This
happens in the initial transient of the solution. To estimate such nonlinearities, observed samples around the peaks of the exponentials are required. At least 2 samples per nonlinear Michaelis-Menten term in a adjustment factor $\alpha_i$ are needed to estimate its two parameters.

\mypar{Summary}
Our analysis shows that the iPBPK-R approach allows us under the following practical conditions to obtain reasonably accurate parameter estimates via optimization:

\begin{inparaenum}[1)]
\item The distance $|y^{+}_{ij} - y^{-}_{ij}|$, needs to be small enough to produce stable parameter estimates.
\item At maximum $n$ \emph{functionally independent} adjustment factors $\alpha_i$ in the linearized matrix can be estimated.
\item At maximum $n$ nonlinearities (2 parameters per nonlinear term) can be estimated.
\item A solution is guaranteed and can be investigated to assess its stability and plausibility.
\item Since iPBPK-R solves a nonlinear high-dimensional optimization problem, there is no guarantees that the global optimum is found.
\item The objective function $\varPsi^{\vec{r},\vec{u}}(.)$ introduces penalty terms to push the optimization towards the biologically most plausible parameter solution. 
\item Good starting values (e.g., values chosen based on IVIVE for biological plausibility) are absolutely essential. 
\end{inparaenum}
The discussion in this section provides a worst-case estimate and bounds.
In particular, a smaller number of modes $\theta_i$ may be relevant and observable in the measurement compartment $j$, and all scaling factors $\alpha_i$ may be acting only on these modes.

\section{The PBPK Model underlying iPBPK-R for EBT}
\label{sec:mtd}
\begin{figure}[t]
\centering
\includegraphics[width=0.6\columnwidth]{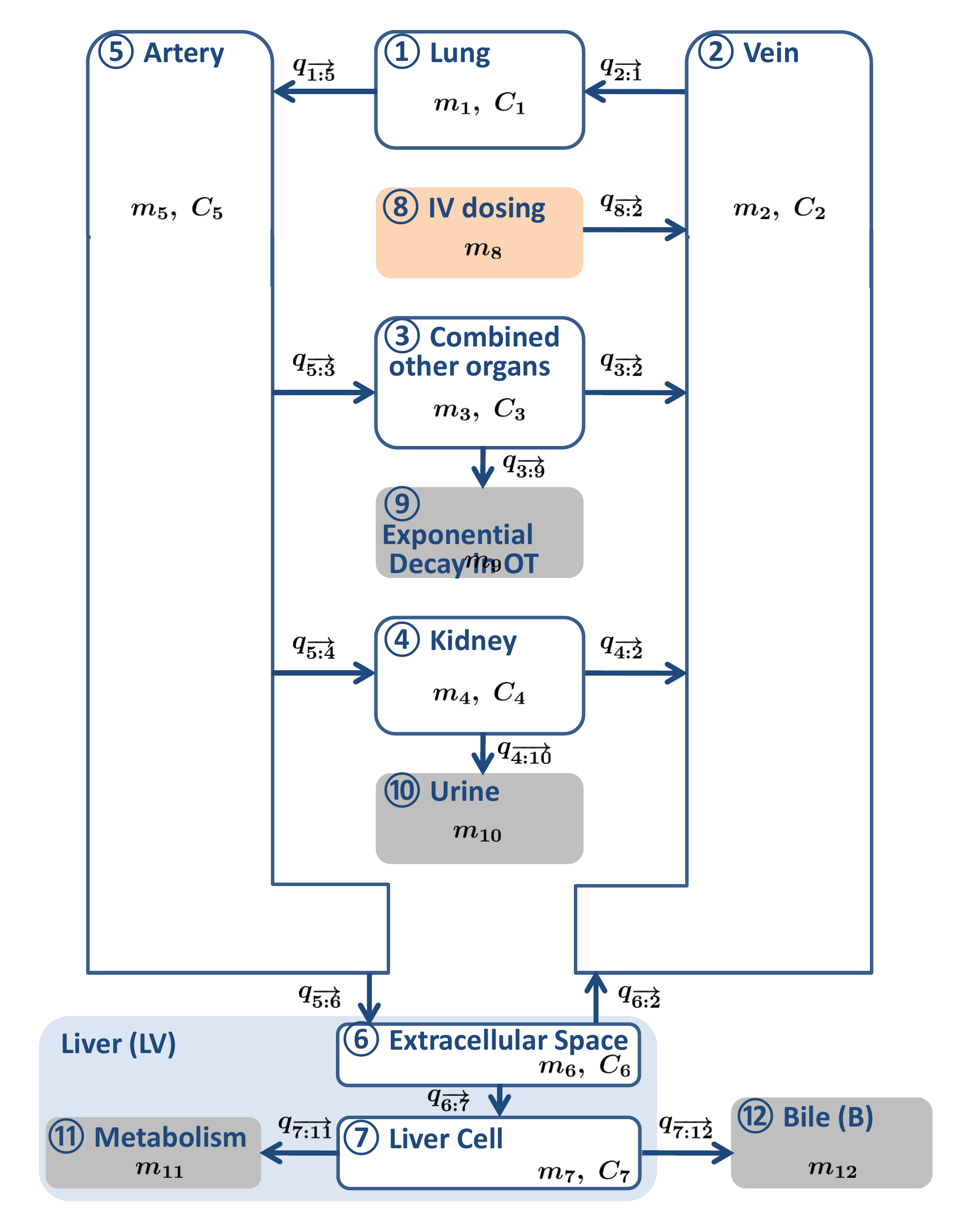}
\caption{The ODE system of $^{14}$C-erythromycin in iPBPK-R.
Adapted from Franchetti et al.~\cite{RN760} with permission of American Society for Pharmacology and Experimental Therapeutics (ASPET).\label{fig:ebtmb1}}\vspace*{-3mm}
\end{figure}

In this section we provide the detailed model description of iPBPK-R as used for the EBT as an instance of the framework laid out in \cref{sec:problem}. We provide the classical view of the PBPK compartment model in \cref{sec:classical} and then translate it into our framework for analysis in \cref{sec:matrixanalysis}. For readability we are collecting all parameters and detailed formulas in \cref{table:2}--\cref{table:8} in the appendix.

\subsection{Developing the Reduced Model}\label{sec:classical}

\begin{figure}[ht]
\centering
\includegraphics[width=0.5\columnwidth]{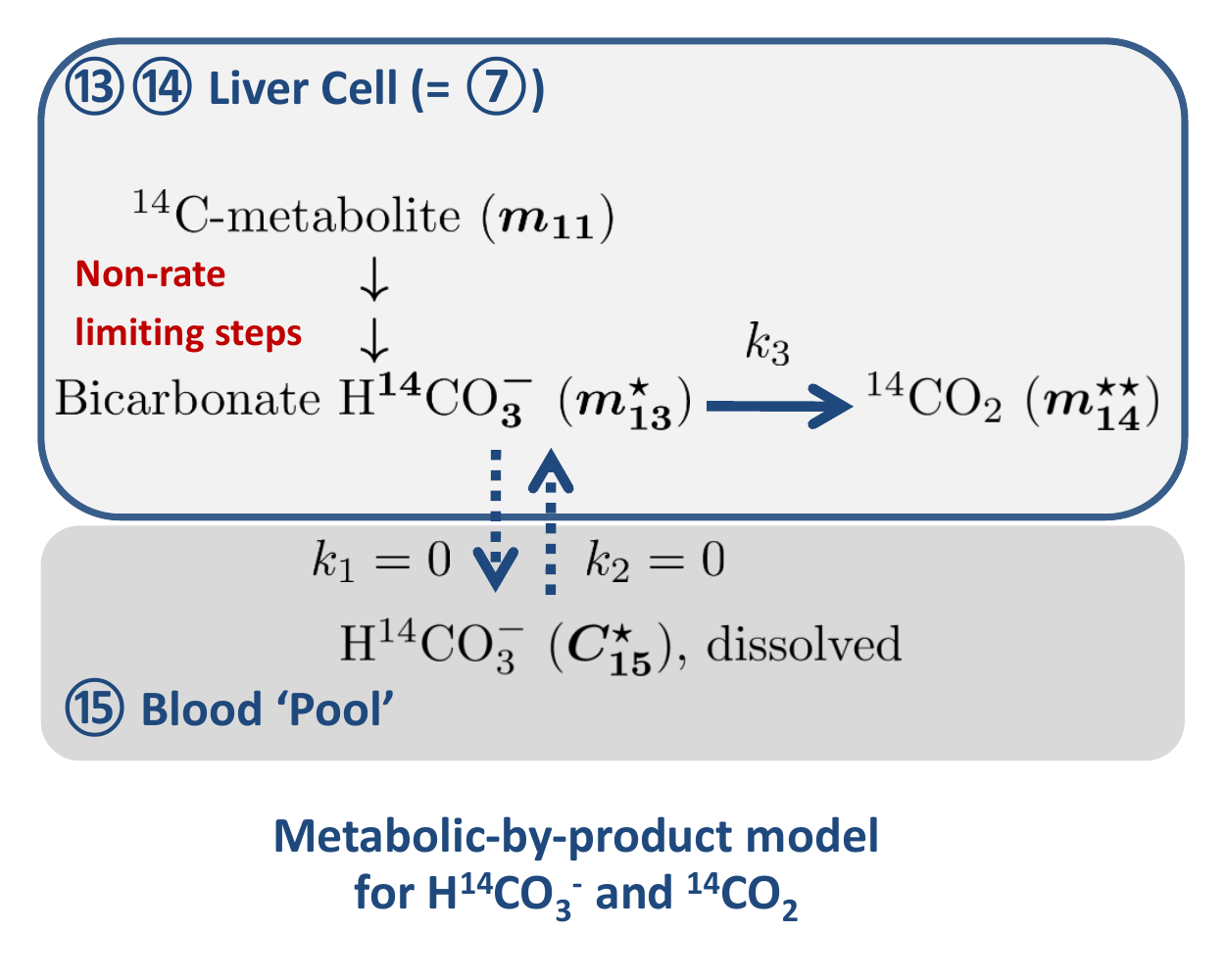}
\caption{The metabolic by-product model in iPBPK-R. \label{fig:ebtmb2}}\vspace*{-3mm}
\end{figure}

\mypar{Model setup}
The core of our system is a seven-compartmental PBPK model of $^{14}$C-erythromycin as shown
Fig.~\ref{fig:ebtmb1}, depicting the ODE system of $^{14}$C-erythromycin. The model contains compartments for artery, vein, lung, kidney, extracellular space, liver cells, and combined other organs. Further, there are supporting source and sink compartments used for mass balance that capture IV dosing, exponential decay in other organs, urine, bile, and metabolic by-product $^{14}$CO$_2$.
Our ODE system is a minimum full model according to
Sager et al. \cite{RN661} but can be also viewed as a reduced order
model since the number of compartments are limited so that parameters in the model are estimable. Building an iPBPK-R model (in our case of $^{14}$CO$_2$ production rate in healthy subjects) entails defining compartments (Fig.~\ref{fig:ebtmb1} and Fig.~\ref{fig:ebtmb2}) and flow terms between the compartments (\cref{table:2}--\cref{table:4}).

\mypar{Mass flows}
The mass flows $q_{ji}(t)$ in \cref{eq:sumq} are expressed as a function of drug concentration $C_i$ and $C_j$ in the compartments $V_i$ and compartment $V_j$, respectively, and parameterized by a vector of adjustment factors,
\begin{equation}\label{eq:rvec}
 \vec{r}=(\alpha_k,\beta_k, \gamma_k,\gamma_i, \gamma_j,\lambda_k,\kappa_{ij})
\end{equation}
  (see \cref{table:4}), which
will be optimized for parameter estimation (see \cref{sec:procedure}).

Mass flows are captured by three types $\mathcal{Q}_1$, $\mathcal{Q}_2$, or $\mathcal{Q}_3$  shown in  \eqref{eq:linearconcflow}--\eqref{eq:diffconcflow} below.
A flow term $q_{ji}(t)$ is given by a single such term or a combination of two or three terms of the mass flows types,
\begin{equation}\label{eq:flowshapes}
q_{ji}^{\vec{r},\vec{u}}(t) 
             = \begin{cases}
              (-1)^{\delta_{ki}} \,\mathcal{Q}_1^{\vec{r}_1}\ \text{or}\ (-1)^{\delta_{ki}} \, \mathcal{Q}_2^{\vec{r}_2}\ \text{or}\ \mathcal{Q}_3^{\vec{r}_3} \\
                (-1)^{\delta_{ki}} \,(\mathcal{Q}_1^{\vec{r}_1} + \mathcal{Q}_2^{\vec{r}_2})\\
                (-1)^{\delta_{ki}} \,(\mathcal{Q}_1^{\vec{r}_1} + \mathcal{Q}_2^{\vec{r}_2}) + \mathcal{Q}_3^{\vec{r}_3}.
                \end{cases}
\end{equation}
Here, $\vec{r}_1 = (\alpha_k,\gamma_k)$, $\vec{r}_2 = (\alpha_k,\gamma_k, \beta_k, \lambda_k, h)$, and $\vec{r}_3 = (\kappa_{ij},\gamma_i, \gamma_j)$,
$\vec{u}$ is a configuration vector, $h,\,k \in \vec{u}$ are configuration constants, where $h=0/1$ implies competitive/non-competitive mass flow inhibition and $k=i$ or $j$. $\delta_{mn}$ is the \textit{Kronecker delta} function (0 for $m \neq n$ and 1 otherwise).

\textit{$\mathcal{Q}_1$ is linearly proportional to the drug concentration $C_k$} of a single compartment,
\begin{equation}\label{eq:linearconcflow}
\mathcal{Q}_1^{(\alpha_k,\gamma_k)} = \alpha_k \mathrm{K}_{k,1}\frac{\mathrm{L}_{k}}{\gamma_k \mathrm{P}_{k}} C_k, \ k=i\ \mathrm{or}\ j,
\end{equation}
with IVIVE constants $\mathrm{K}_{k,1}$ (\textrm{clearance}) 
and $\mathrm{L}_{k}$ (captures fraction unbound).
\textit{$\mathcal{Q}_2$ is nonlinearly proportional to the drug concentration $C_k$} of a single compartment,
\begin{equation}\label{eq:nonlinearconcflow}
\mathcal{Q}_2^{(\alpha_k,\gamma_k, \beta_k, \lambda_k,h)} = \frac{\beta_k^{\delta_{0h}} \mathrm{K}_{k,21}\frac{\mathrm{L}_{k}}{\gamma_k \mathrm{P}_{k}} C_k}{\lambda_k^{\delta_{1h}} \mathrm{K}_{k,22} + \frac{\mathrm{L}_{k}}{\gamma_k \mathrm{P}_{k}} C_k}, \ k=i\ \mathrm{or}\ j,
\end{equation}
with IVIVE constants $\mathrm{K}_{k,21}$ (maximum drug mass flow), and $\mathrm{K}_{k,22}$ (drug concentration where 50\% of $\mathrm{K}_{k,21}$ is achieved).
\textit{$\mathcal{Q}_3$ is proportional to the difference in drug concentration} between compartments $C_i$ and $C_j$,
\begin{equation}\label{eq:diffconcflow}
\mathcal{Q}_3^{(\kappa_{ij},\gamma_i, \gamma_j)} = \kappa_{ij}\Big(\mathrm{K}_{j,3} \frac{\mathrm{L}_{j}}{\gamma_j \mathrm{P}_{j}} C_j - \mathrm{K}_{i,3} \frac{\mathrm{L}_{i}}{\gamma_i \mathrm{P}_{i}} C_i\Big),
\end{equation}
where $\mathrm{K}_{i,3}$ and $\mathrm{K}_{j,3}$ combine multiple IVIVE values. 

\mypar{Vein and artery compartments}
In iPBPK-R, Vein and Artery compartments were separately modeled. 
This differs from classical PBPK models where in clinical pharmacology blood tissue is viewed as one compartment. In the connection from Artery, Extracellular Space, through to Vein compartment, we assumed the two algebraic relationships (see Fig.~\ref{fig:ebtmb1}). First, the mass flow from Artery to Extracellular Space $q_{\overrightarrow{\Art:\ES}}$ is fast and thus modeled instantaneous as
$
            C_{\Art|\ES}=C_{\ES}
$. 
Second, we assumed the 2-blood compartment model (i.e., a model with Vein and Artery compartments) for the blood tissue to be well-stirred, which necessitated a second algebraic relationship
\begin{equation}\nonumber
C_{\VN|\ES} = \frac{C_{\ES}}{\PTES}.
\end{equation}
Both are necessary assumptions in iPBPK-R to approximate the drug flows between
the blood tissue and liver organ when the blood is modeled with two separate compartments,
Artery and Vein, while retaining the well-stirred model assumption for the blood tissue.

\mypar{Modeling $^{\boldsymbol{14}}$CO$_{\boldsymbol{2}}$ dynamics}
Beyond the drug and its metabolite, $^{14}$CO$_2$ production rate in the EBT needs to be modeled, which we conduct via a one-compartment metabolic-by-product model following \cite{RN622} 
(Fig.~\ref{fig:ebtmb2}).
Given the separation of time scales between the drug dynamics and the $^{14}$CO$_2$ dynamics, we model $^{14}$C-erythromycin as converted to its metabolite and by-product $\mathrm{H^{14}CO}^-_3$ in the CYP3A4-mediated pathway instantaneously and the subsequent final-by-product $^{14}$CO$_2$ transported instantaneously to the lung and exhaled.
This implies
the two rate constants $k_1$ and $k_2$ from \cite{RN622} are set to 0, which is supported by an in vivo rat study.
Thus,  $^{14}$CO$_2$ gets released from $\mathrm{H^{14}CO}^-_3$ in the liver cell with an excretion rate constant $k_3$ and is instantaneously measured in the breath.
$C^{\star}_{\LCBC}$ and $C^{\star}_{\BLPO}$ denotes the concentration of $\mathrm{H^{14}CO}^-_3$ in the liver cell and in the blood pool. The respired amount of $\mathrm{H^{14}CO}^-_3$ and exhaled $\mathrm{^{14}CO}_2$ have a one-to-one molar relationship. 
The amount of $\mathrm{^{14}CO}_2$ in the breath is denoted by $m^{\star\star}_{\LCCD}$ (see \cite{RN760}).

\subsection{Matrix Representation of the PBPK Model}\label{sec:matrixanalysis}
In this section we present the reduced model used in iPBPK-R in the formalism described in  \cref{subsec:matrixODEs}.
The full non-linear system of ODEs is written as
\begin{equation}\label{eq:odeconcEBT}
\frac{d\vec{m}(t)}{dt} = \mathbf{X}\vec{m}(t) + \mathbf{\tilde{Y}}(\mathbf{D}\vec{m}(t)) + z(t).
\end{equation}
Non-zero entries $x_{i,j}$ of $\mathbf{X}$, $y_{i,j}(.)$ of $\mathbf{\tilde{Y}}$, and $d_{i}$ of the diagonal matrix $\mathbf{D}$ are provided in \cref{table:8}.
The diagonal elements $d_i$
pre-scale the vector elements $m_i(t)$ before $\mathbf{\tilde{Y}}(.)$ is applied and
 can be propagated into the parameters of the $y_{i,j}(.)$ using
the identity
\begin{equation}\nonumber
\mathrm{MM}_{a,b,c}(\alpha x) = \mathrm{MM}_{\alpha a,b,\alpha c}(x).
\end{equation}
This allows us
to convert \eqref{eq:odeconcEBT} into the normal form given by \eqref{eq:odeconc} with
system operator matrix
\begin{equation}\nonumber
\mathbf{A}(.) = \mathbf{X}(.) +
\mathbf{Y}(.)
\quad\text{with}\quad
\mathbf{Y}(.)
 = \mathbf{\tilde{Y}}(.) \circ \mathbf{D}(.).
\end{equation}

The forcing function $z(t)$ follows the shape discussed in \cref{sec:problem}. It models injecting the drug at a dosing rate $\tau=M_0/t_0$ for the short dosing interval $\Omega = [0,\,t_0]$ into the Vein compartment.
In \eqref{eq:odeconcEBT} and \cref{table:8} we omit the bicarbonate concentration in the pool compartment, $C^{\star}_{\BLPO}$, since $k_1 = k_2 = 0$ was assumed (\cref{table:2}). As a result, the total number of ODEs in iPBPK-R of EBT for healthy subjects is $n=14$.

\mypar{Analysis}
Having expressed iPBPK-R's reduced PBPK model and all its dependencies on parameters and adjustment factors as instance of \eqref{eq:odeconc} enables us to analyze iPBPK-R's estimation capabilities and stability.
We want to emphasize that iPBPK-R uses standard numerical solvers for ODE integration and optimization. The theoretical framework developed in \cref{subsec:accuracy} is \emph{not} used to compute the solutions but to analyze the solution quality and soundness of the approach.

We base the discussion below on analyzing the eigendecompositions of the upper and lower matrix approximations $\mathbf{A}^+$ and $\mathbf{A}^-$ of the system operator matrix $\mathbf{A}$ of \eqref{eq:odeconc}. The element-wise distance
$|a^+_{i,j}-a^-_{i,j}|$ is small as required, and both $\mathbf{A}^+$ and $\mathbf{A}^-$ are indeed simple matrices.

The first observation is that while the EBT iPBPK-R model has 14 compartments for the purpose of modeling mass balance, the underlying dynamical system has only 7 compartments. The 5 source/sink compartments and the 2 further helper compartments lead to a null space of dimension 6 and can be disregarded in the discussion.
Inspection of Fig.~\ref{fig:twelvefit}  and the simulated curves in all 7 compartments (shown in \cite{RN760}) as well as Fig.~\ref{fig:twofit} indicate
that at most four distinctive estimable modes $\theta_i$ of the eigendecomposition are present in the liver compartment.  Thus, the $T=11$ sampling points and the associated noise is sufficient to estimate the modes to a reasonable accuracy level. The measurements are clustered at the early transient phase and thus allow for estimating non-linear saturation due to Michaelis-Menten behavior.

Secondly, the number of adjustment parameters that can be distinguished is limited to $M=7$ functionally independent $\alpha_i$ at max. iPBPK-R has $\mathrm{K}=9$ adjustment parameters, and thus they must have a functional (non-linear) dependence. The dependence is actually shown in \cite{RN760} as the four main levers that shape the $^{14}$CO$_2$ production rate curve. The condition that the adjustment factor vector $\vec{r}$ needs to be as close to $\mathbf{1}_M$ as possible disambiguates the functional dependency. This regularizes the estimates and
pushes them towards the global optimum analogous to the arguments presented in \cite{RN872}.
Adding to estimation stability is the nested optimization procedure employed by iPBPK-R, presented next.

\section{Parameter Estimation in iPBPK-R}
\label{sec:procedure}
\mypar{Overview}
Multiple-parameter estimation via iPBPK-R is 
an optimization problem \eqref{eq:opt} for a subset of parameters in the system of ODEs in \cref{eq:sumq},
collected in a $K$-dimensional parameter vector $\vec{r}$ 
as defined in \eqref{eq:rvec}.
This optimizes a subset of drug flow terms (see \eqref{eq:flowshapes} and \cref{table:3}) by minimizing the distance 
between observed and simulated data.
Some of the optimized adjustment factors will provide parameter estimates for activities of metabolic enzyme and drug transporters in individuals, in non-renal drug clearance, 
providing an indirect measurement procedure for these parameters. 
We developed an EBT specific objective function (an instance of \cref{eq:objf}) consisting of a distance measure and  
penalty terms that regularizes the problem. 

\mypar{$^{\boldsymbol{14}}$CO$_{\boldsymbol{2}}$ measurement}
Remember from \cref{subsec:Ltwonorm} that $B(t)$ denotes the $^{14}$CO$_2$ production rate at time $t$ and is a function $\mathcal{F}(.)$ of the solution vector $\vec{C}(t)$.
Further, $\dot{C}_{i^{\star}}(t)$ denotes the
derivative of concentration in the $i^{\star}$ compartment. In our case this is the Liver compartment, which metabolizes the drug into $^{14}$CO$_2$.
Thus,
\begin{equation}\nonumber
B(t) = \mathcal{F}(\vec{C}(t)) = \dot{C}_{i^{\star}}(t)
\end{equation}
describes $^{14}$CO$_2$ production rate measured via breath and is an instance of \eqref{eq:breath}.

\mypar{Configuration} Next we define the constant configuration vector
\begin{equation}\nonumber
\vec{u} = \big(h, \vec{h}', \vec{h}'', (u_{\ell})_\ell, u_{\mathrm{b}}, u_{\mathrm{c}}, u_{\mathrm{d}}, u_{\mathrm{x}}\big)
\end{equation}
with a scalar $h$, $K$-dimensional vectors $\vec{h}'$ and $\vec{h}''$, and weights $u_{\ell}$. $h$ is the indicator used in \cref{eq:nonlinearconcflow} while
$\vec{h}'$ and $\vec{h}''$ 
configure penalties for adjustment factors in $\vec{r}$. $(u_{\ell})_{\ell}$ is weighting the measurement while $u_{\mathrm{b}}$, $u_{\mathrm{c}}$, $u_{\mathrm{d}}$, and $u_{\mathrm{x}}$ are penalty weights in \cref{eq:pi}.

\begin{figure}[t!]
\centering
\includegraphics[width=0.8\columnwidth]{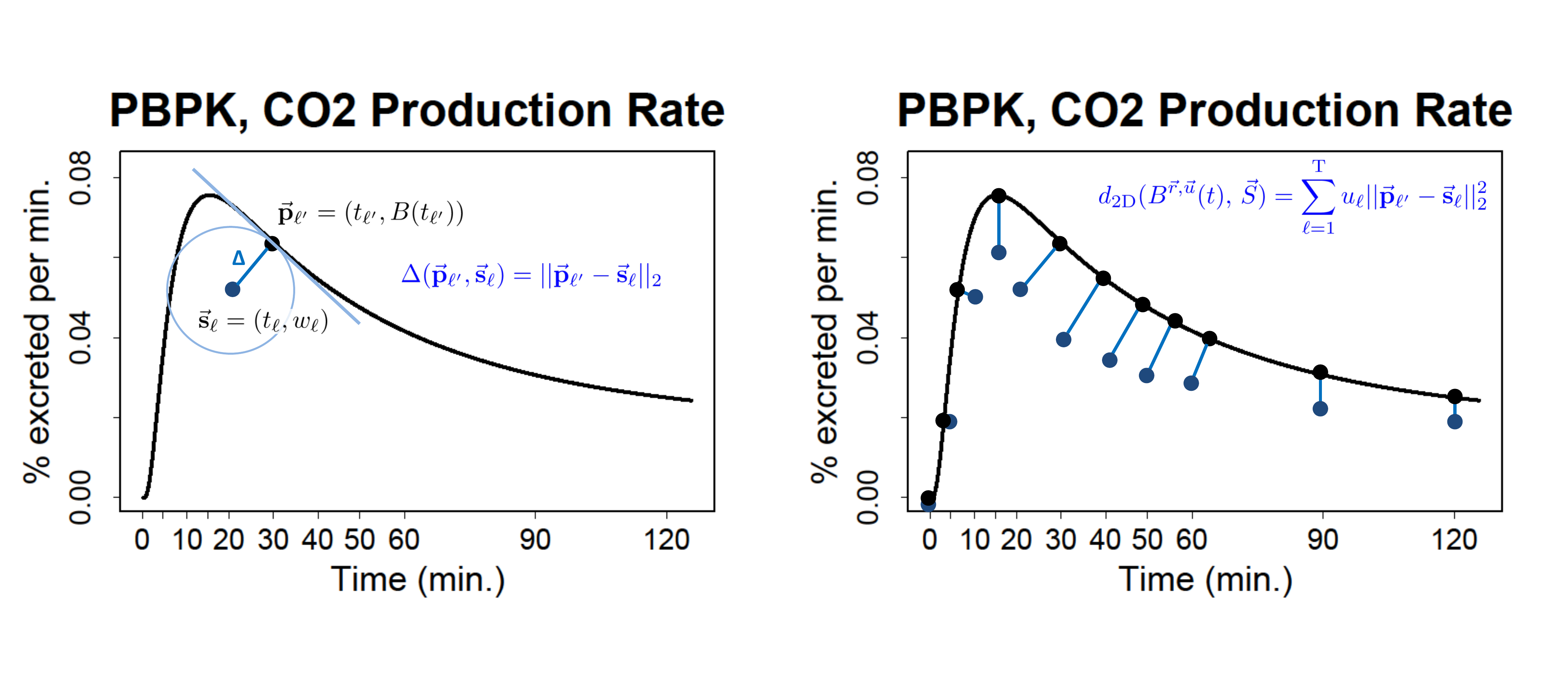}
\caption{2D L$_2$ norm and 2D distance measure\label{fig:twodnorm}.}
\vspace*{-5mm}
\end{figure}

\mypar{Distance measure}
Clinical measurements may be slightly shifted relative to the planned time point.
We introduce a 2D distance (Fig.~\ref{fig:twodnorm}) between 
simulated and measured $^{14}$CO$_2$ data points that
allows some (penalized) misalignment in $x$ direction in addition to mismatch in $y$ direction.

We define the $\ell$th measurement point measured at time $t_\ell$ as $\vec{s}_\ell = (t_\ell, w_\ell)$ and
the point $p_{\ell'} = (t_{\ell'}, B(t_{\ell'}))$
where
\begin{equation*}
t_{\ell'} =
\underset{t' \in [t_\ell-\delta, t_\ell+\delta] }{\arg\min} \|(t', B(t')) - (t_\ell, w_\ell)\|_2
\end{equation*}
for an appropriately chosen $\delta$. We
collect the $T$ measurements $\vec{s}_\ell$  at time points $t_\ell$ as vector
$\vec{S} = \big((t_\ell, \vec{s}_{\ell})\big)_{\ell}$.
Then the 2D distance between the simulated function $B(t)$ parameterized by $\vec{r}$ and $\vec{u}$ and the measurement vector $\vec{S}$,
is defined as
\begin{equation}\label{eq:twodnormd}
d_{\mathrm{2D}}(B^{\vec{r},\vec{u}}(t),\,\vec{S})  
 = \sum_{\ell=1}^{\mathrm{T}} u_{\ell}||\vec{p}_{\ell'} - \vec{s}_{\ell}||^2_2.
\end{equation}

\mypar{Objective function in iPBPK-R}
We now state the full objective function as used for EBT (an instance of \eqref{eq:objf}),
\begin{equation}\label{eq:dtwodobjc}
\varPsi^{\vec{r},\vec{u}}_{\mathrm{2D}}(B^{\vec{r},\vec{u}}(t),\,\vec{S}) =
  d_{\mathrm{2D}}(B^{\vec{r},\vec{u}}(t)),\,\vec{S}) +
  \pi(\vec{r}, \vec{u})
\end{equation}
with the components of the penalty and regularization term $\pi(\vec{r}, \vec{u})$ defined in
 \cref{table:OVF}, instantiating \eqref{eq:pi}:
\cref{eq:bias} avoids biologically unrealistic parameter estimation by penalizing distance from IVIVE values,
\cref{eq:lbound} ensures non-negative estimates,
\cref{eq:drift} penalizes peak height deviation from the data peak, and
 \cref{eq:xshift} penalizes mis-alignment in the $x$ direction.
Variants of $\varPsi^{\vec{r},\vec{u}}_{\mathrm{2D}}(.,.)$ adjusted to problem settings can be configured via the configuration vector $\vec{u}$.

\mypar{Parameter estimation}
Biologically plausible multiple-parameter estimation via iPBPK-R for measured samples $\vec{S}$ is done by solving the optimization problem
\begin{equation}\label{eq:optimtailoredtwod}
\vec{r}
= \underset{\vec{r}'}{\arg\min}\, \varPsi^{\vec{r}',\vec{u}}_{\mathrm{2D}}\big(\mathcal{F},\,\vec{S}\big)
\quad\text{for configuration\ }\vec{u}
\end{equation}
to estimate the most likely parameter vector $\vec{r}$. The full optimization space for EBT allows for 7 independent parameters, but in practice not all of them are strongly influencing the fit. Depending on the goodness-of-fit of modeling, some less influential parameters in $\vec{r}$ can be fixed as constants and not necessarily be optimized. This is achieved through appropriate setting of the indicator vectors $\vec{h}'$ and $\vec{h}''$.

\mypar{Combined pre/post treatment estimates}\label{subsec:cooptimization}
The focus of clinical pharmacology studies often is the impact of a treatment (in our case dialysis) on biological parameters.
In a kidney disease study where EBT is used before and after the patients receive dialysis, our goal is to apply iPBPK-R to these two data series to estimate the effect of dialysis on biological parameters of individual patients.
We assume that \textit{independent parameters} remain unchanged while \textit{co-optimization parameters} change from pre-dialysis to post-dialysis.
To estimate both co-optimization parameters and independent parameters we developed a co-optimization method as discussed below.

The same data sampling procedure using configuration $\vec{u}$ is performed pre and post dialysis, leading to data sets
$\vec{S}_1$ and $\vec{S}_2$.
From the corresponding parameter estimates $\vec{r}_1$ and $\vec{r}_2$ we derive the simulated breath curves $B_1(t)$ and $B_2(t)$. The parameters vectors $\vec{r}_1$ and $\vec{r}_2$ are not independent:  we set
$$
\vec{r}_1 = \vec{r}_\mathrm{c} \oplus \vec{r}_{\mathrm{i},1}
\quad\text{and}\quad
\vec{r}_2 = \vec{r}_\mathrm{c} \oplus \vec{r}_{\mathrm{i},2}
$$
where $\vec{r}_\mathrm{c}$ denotes the co-optimized parameters and $\vec{r}_{\mathrm{i},j}$ the independent parameters, and
$\oplus$ denotes the vector direct sum.
With this setup we define the co-optimization problem as
\begin{equation}\label{eq:cooptimtailoredtwod}
(\vec{r}_1, \vec{r}_2)
= \underset{\vec{r}'_\mathrm{c}, \vec{r}'_{\mathrm{i},1}, \vec{r}'_{\mathrm{i},2}}{\arg\min}\,
\sum_{j=1}^2
\varPsi^{\vec{r}'_\mathrm{c} \oplus \vec{r}'_{\mathrm{i},j},\vec{u}}_{\mathrm{2D}}\big(B_j^{\vec{r}'_\mathrm{c} \oplus \vec{r}'_{\mathrm{i},j},\vec{u}}(t),\,\vec{S}_j\big).
\end{equation}
Parameters beyond $\vec{r}$ defined in  \eqref{eq:rvec} may need to be included in  $\vec{r}_1$ and $\vec{r}_2$, and $\vec{u}$ needs to be chosen carefully to
model pre/post dependencies and address particular research.

\mypar{Numerical methods and implementation}
We implemented the iPBPK-R in the R system. We used the R function \texttt{ode} in the \texttt{deSolve} package that implements a suite of ODE solvers, including explicit and implicit solvers, adaptive solvers, and Runge Kutta solvers \cite{RN752} \cite{RN754}. Optimization was performed using the \texttt{L-BFGS-B} and \texttt{BFGS} algorithms provided by the R function \texttt{optim} that implements  quasi-Newton method with or without a limited-memory modification, respectively \cite{RN753}. 
The nested optimization used for co-optimizing estimates is implemented via a generalized Hill Climbing approach. Optimization of the independent parameter vectors $\vec{r}_{\mathrm{i},1}$ and $\vec{r}_{\mathrm{i},2}$ were performed first independently in a inner loop. Optimization of the co-optimization parameter vector $\vec{r}_\mathrm{c}$ was implemented in an outer loop. This iteration was performed iteratively until sufficient convergence.

\begin{table}
\centering
\abovedisplayskip=0pt
\belowdisplayskip=0pt
\begin{tabular}{m{2cm} m{10cm} }
\hline 
Bias & {\begin{flalign}
\nu(\vec{r}, \vec{h}')   = \| \mathrm{diag}(\vec{h}')  (\vec{r} - \vec{r}_0)\|^2_2
\quad \mathrm{with}\quad  \vec{h}' \in\{0,1\}^{\mathrm{K}}&&
\label{eq:bias}\end{flalign}}\\[-2.4ex]
Lower bound & {\begin{flalign}
\rho(\vec{r}, \vec{h}'')  =
\sum_{i=1}^{\mathrm{K}} h''_i \min(r_i - r_i^\mathrm{L}, 0)^2
\quad \mathrm{with}\ 
\vec{h}'' \in\{0,1\}^{\mathrm{K}}&&
\label{eq:lbound}\end{flalign}}\\[-2.4ex]
Drift & {\begin{flalign}
            \epsilon(\vec{r})  = \Big(\max_{t \in [t_0,\,t_{\mathrm{T}}]}{B^{\vec{r},\vec{u}}(t)} - \max_{\ell=1,\dots,\mathrm{T}}{w_{\ell}}\Big)^2 &&
\label{eq:drift}\end{flalign}}\\[-2.4ex]
X-shift & {\begin{flalign}
\psi(\vec{r}) = \sum_{\ell=1}^{\mathrm{T}}|t_{\ell'} - t_{\ell}|&&
\label{eq:xshift}\end{flalign}}\\[-2.4ex]
\\
\hline
\end{tabular}
\vspace*{-1mm}\caption{Regularization penalty terms.}\label{table:OVF}
\end{table}

\section{Examples and Results}
\label{sec:experiments}
This section  provides a brief summary of two applications of iPBPK-R and discusses the results with respect to method feasibility and soundness.
The detailed results and pharmacological interpretations of the first application had been described in \cite{RN760}. In addition, the preliminary results of the second application was presented as a poster \cite{RN761}.

\subsection{Estimating Parameters in Healthy People}\label{subsec:exampleHealthy}
We model $^{14}$CO$_2$ production rate data obtained from 12 healthy subjects in an EBT study to estimate biological parameters. The study design is described in \cref{subsec:ebt} and \cite{RN728}. Briefly, 12 healthy subjects received a single IV dose of $^{14}$C-erythromycin (0.074 mmol), and breath samples were collected immediately at 11 time points within two hours including the baseline time point of EBT. $^{14}$CO$_2$ production rates in the collected breath samples were calculated.

The iPBPK-R model structure and parameters of the system of ODEs were shown Fig.~\ref{fig:ebtmb1}, Fig.~\ref{fig:ebtmb2}, and \cref{table:4}.
The initial values of the input parameters are shown in \cref{table:2}, including IVIVE values can be found in \cite{RN760}. Through an initial manual investigation we first identified that nine parameters were either impacting the shape of the simulated rate-time curve $B(t)$ or biologically relevant parameters (see \cref{table:4})
\begin{equation*}
\VmaxupES,\ \VmaxpgpLC,\ \VmaxmrpLC,\ \CLcyp,\ \VLC,\ \PTOT,\ \mu_{\OT},\ \QOT,\ \text{and}\ \QESLC,
\end{equation*}
and their respective adjustment factors to be optimized,
\begin{equation*} 
\beta_{\ES},\ \beta_{\LC,1},\ \beta_{\LC,2},\ \alpha_{\LC},\ \zeta_{\LC},\ \gamma_{\OT},\ \alpha_{\OT},\ \kappa_{\OT\Art},\ \text{and}\
\kappa_{\ES\LC}.
\end{equation*}

\begin{figure}[t!]
\vspace*{-3mm}
\centering
\includegraphics[width=\columnwidth]{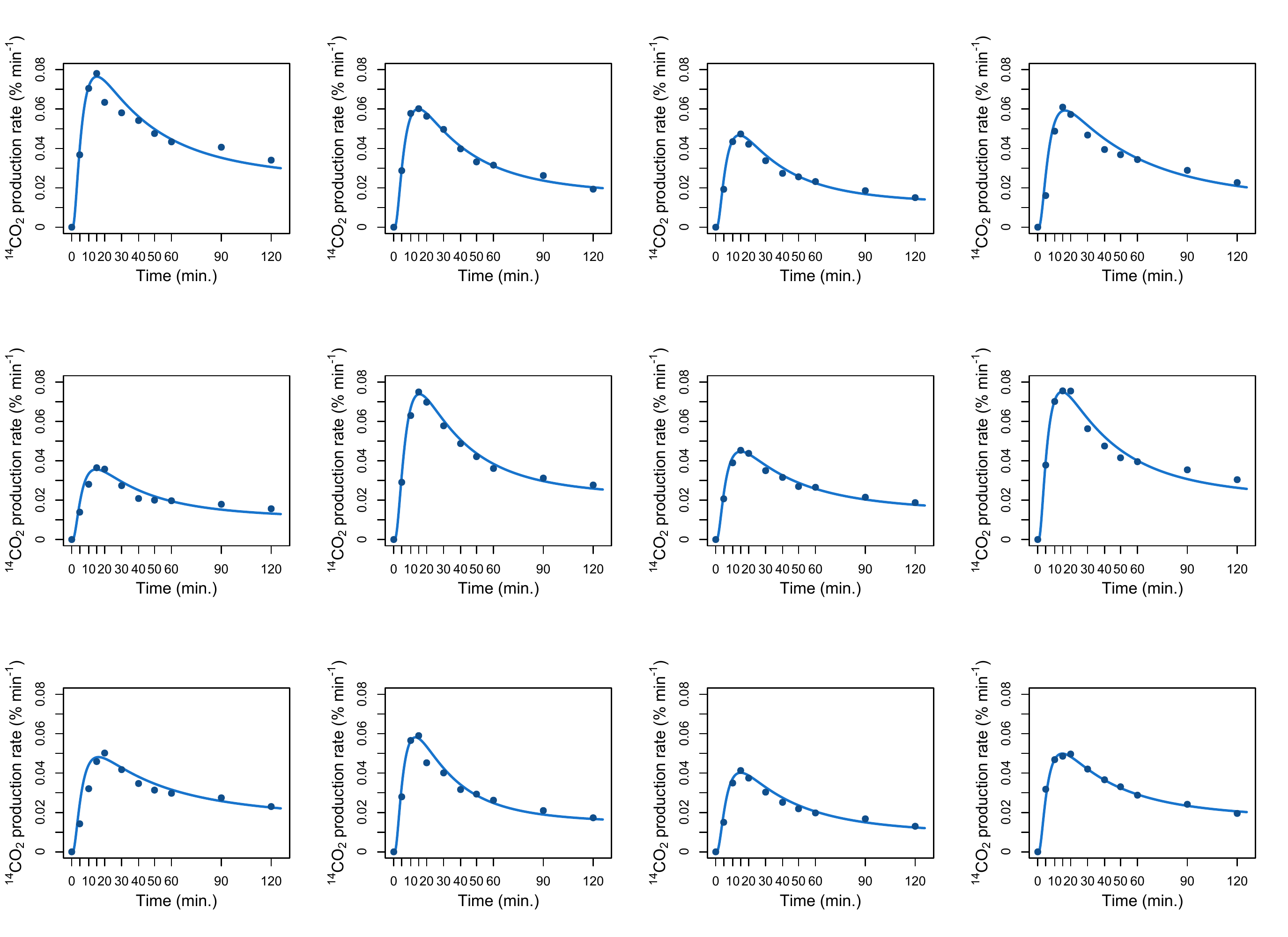}
\caption{iPBPK-R model fit to the $^{14}$CO$_2$ production rate-time curves of 12 healthy subjects.
Adapted from Franchetti et al. \cite{RN760} with permission of ASPET.\label{fig:twelvefit}}
\end{figure}

\begin{figure}
\centering
\includegraphics[width=0.6\columnwidth]{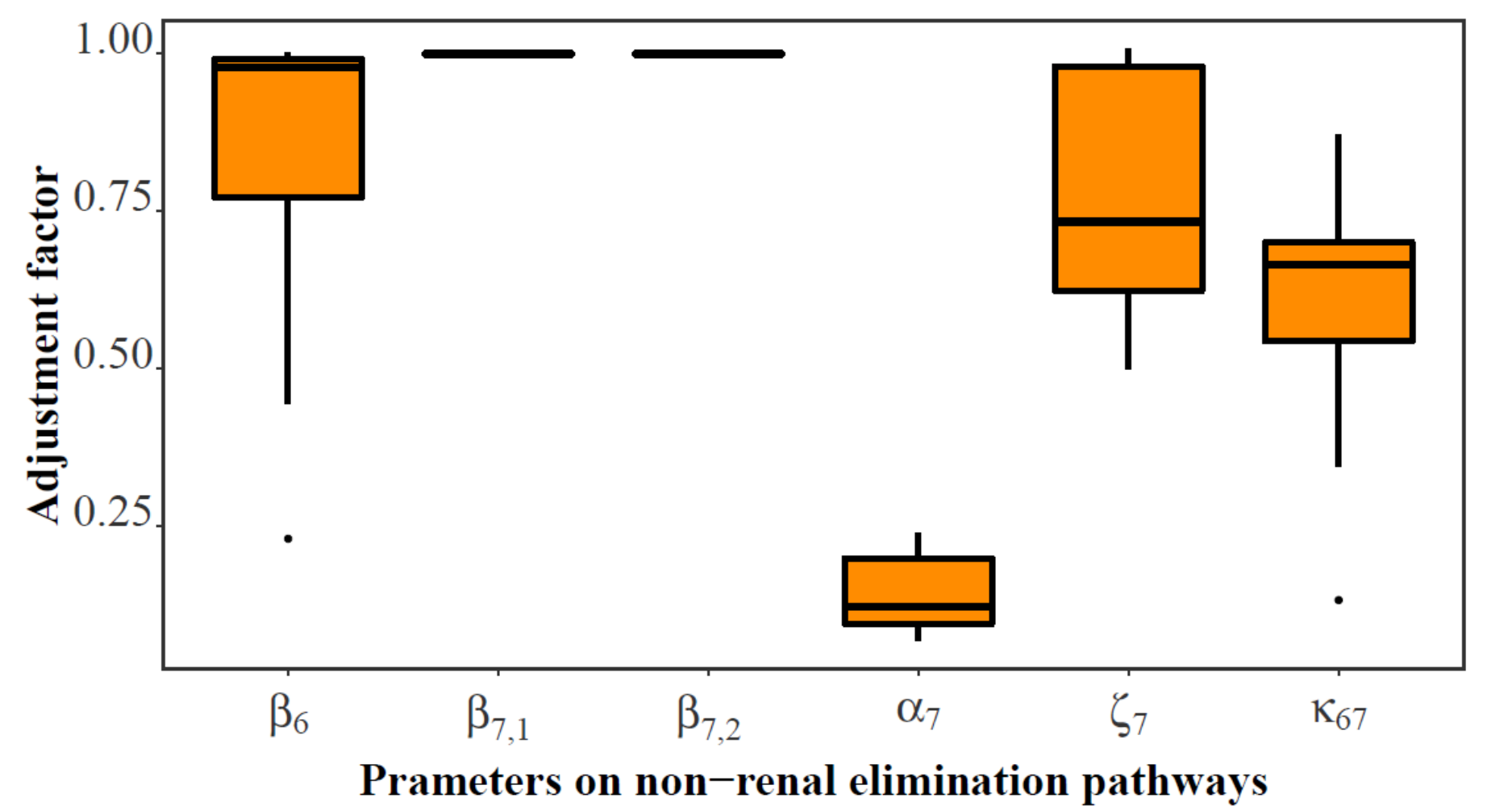}
\caption{Box plot of estimated adjustment factors in 12 subjects.
Adapted from  \cite{RN760} with permission of ASPET. \label{fig:twelveest}}
\vspace*{-5mm}\end{figure}

The results are summarized in Fig.~\ref{fig:twelvefit}. The iPBPK-R model fit well for individual $^{14}$CO$_2$ production rate-time curves
as result of parameter optimization. Sensitivity analysis and analysis following \cref{subsec:accuracy} established that
the parameter estimates are sound. Among the optimized adjustment factors $\beta_{\ES}$, $\beta_{\LC,1}$, $\beta_{\LC,2}$, $\alpha_{\LC}$, $\zeta_{\LC}$, and $\kappa_{\ES\LC}$
were associated with the Liver compartment, where activities in non-renal elimination pathways were target of the published research. The parameter estimates are shown in Fig.~\ref{fig:twelveest}. While 
$\beta_{\LC,1}$ and $\beta_{\LC,2}$ did not have impact on the model fit in the figure, we forced them into the iPBPK-R model for mechanistic considerations.

In exploring the simulation results, the estimated $\alpha_{\LC}$ were stratified by gender. Analysis led to the finding of gender difference in enzyme activity of the liver as reported in \cite{RN760}. This aligned with literature describing a gender difference in expression of the same enzyme in human liver samples \cite{RN639}, a fact not actively modeled in iPBPK-R. While this finding is well beyond the scope of method evaluation, similar evidence will help us to build confidence in steps towards clinical utility.
Production simulation runs required about 10 CPU hours per subject on the \textit{Bridges} supercomputer at PSC.

\subsection{Co-Estimation Pre- and Post Dialysis}\label{subsec:exampleKD}
The second example is the application of iPBPK-R to the $^{14}$CO$_2$ production rate data obtained from 12 patients with kidney disease in an EBT study. The study design of the EBT study is described in \cite{RN19} and \cite{RN761}. Briefly, 12 patients 
received a single IV dose of $^{14}$C-erythromycin before taking a 4-hour dialysis. Two hours post-dialysis these patients received another single IV dose. 
In each EBT breath samples were collected immediately at 11 time points within two hours of IV, including the baseline time point of EBT. Subsequently $^{14}$CO$_2$ production rates 
were calculated for the pre- and post-dialysis EBTs. 
The model structure 
and the system of ODEs were similar to those in the first example, 
and it was assumed that the reduced activity of mass flow is non-competitive. 
The mass flow from Kidney to Urine (see Fig.~\ref{fig:ebtmb1}) was removed assuming 
no kidney function in patients and
drug removal 
by dialysis was assumed negligible.
Further, a pre/post carry-over effect was added that carried the trailing drug concentration from the pre-EBT into the post-EBT.

The same optimization parameters were selected as in the first example. 
Optimization was implemented via nested co-optimization and the adjustment factor
$\alpha_{\LC}$ was optimized in the inner loop to independently estimate
the adjustment factors pre- and post-dialysis for linear CYP3A4 activity. Inhibition parameters associated with drug transporters were also estimated independently so that the change in the nonlinear activity of each drug transporter could be evaluated by comparing the inhibition parameters pre- and post-dialysis.

\begin{figure}[t!]
\vspace*{-3mm}
\centering
\includegraphics[width=0.9\columnwidth]{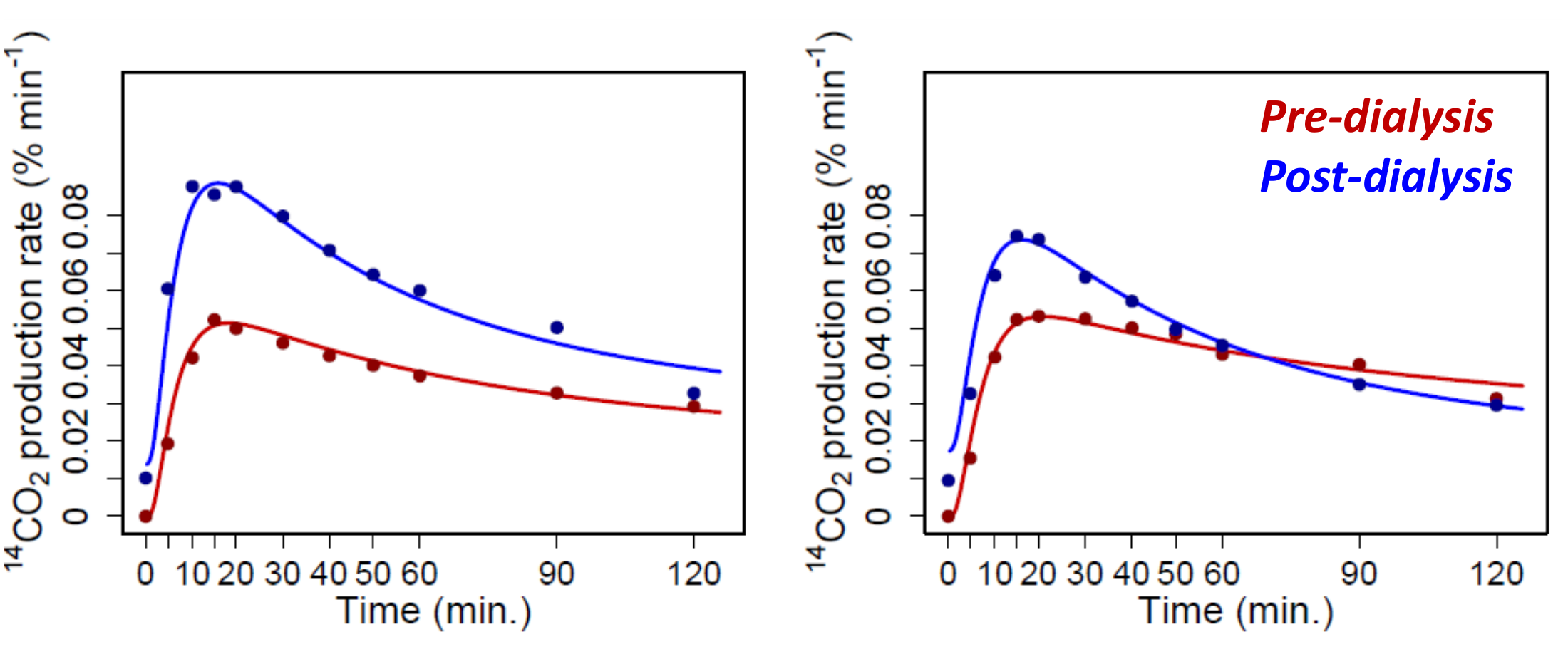}
\caption{iPBPK-R model fit to the $^{14}$CO$_2$ production rate-time curves of two patients with dialysis.\label{fig:twofit}}
\vspace*{-3mm}
\end{figure}

We show the resulting iPBPK-R model fit for two patients 
in Fig.~\ref{fig:twofit}. The iPBPK-R model fit the individual $^{14}$CO$_2$ production rate-time curves pre- and post-dialysis well.
Again, sensitivity analysis and analysis following \cref{subsec:accuracy} establishes that
the parameter estimates are sound.
In further analysis shown in \cite{RN761}, activities of CYP3A4 and drug transporter activities were compared across 
dialysis in each patient so that the effect of dialysis on the non-renal elimination pathways could be evaluated.
A production simulation run required 24 CPU hours per patient. 

\section{Discussion}

\label{sec:discuss}
\subsection{iPBPK-R Summary}
\mypar{Pharmaceutical application}
\textit{iPBPK-R} allows us to estimate multiple physiological parameters in individuals using a single dose of a single probe drug. 
This 
mechanistic indirect measurement approach allows for  \textit{in vivo} activities (e.g., of CYP3A4 and drug transporters) to be \textit{estimated for a particular individual}.  
Activities can be estimated pre/post intervention or disease progression for longitudinal 
monitoring.  
With iPBPK-R mechanistic sources of inter-individual variability can be 
identified, and 
we anticipate it 
to be eventually useful to enable personalized dosing of narrow therapeutic drugs based on 
the estimated physiological activity.

The focus of this paper is to establish iPBPK-R as a method and analyze its properties and performance guarantees for future uses in  emerging breath biopsy/biomarker research \cite{RN1000,RN1001}.
To establish iPBPK-R as an clinically aiding tool, well-designed large-scale clinical trials need to be conducted in a step-wise manner in a probe development context. Such an undertaking will require resources beyond our small group, but will be aided by the foundations laid by this paper. 
Potential future fields of application 
include oncology, pediatrics, nephrology, and severe health conditions. 

\mypar{Key insight}
Our approach uses \textit{production rate} data modeled as the first derivative of drug concentration data in model fitting.
Through the dynamic changes in the early (transient) phase of the 
production rate-time curve, parameters of both rate-limiting and non-rate-limiting steps can be estimated. 
As the probe dose is low, detailed sampling of the initial phase of drug concentration can reveal the characteristics of the system, similar to how impulse responses are used to characterize systems of interest in signal processing  \cite{RN684}.
In future work we will investigate ODE-imposed coupling of multiple drug concentration datasets per individual. 

\mypar{Model and estimation}
We used a reduced order model (that still qualifies as full PBPK model \cite{RN661}) in iPBPK-R where the number of compartments was limited so that model parameters 
can be estimable.
Reducing the number of parameters is essential 
since the main purpose of this method is not to predict PK profiles for populations 
but to back estimate multiple physiological parameters of a nonlinear system of ODEs in particular individuals.
We paid particular attention at how to regularize the optimization problem to 
enable convergence without skewing the estimation results. 

\mypar{Use of biological knowledge as reference}
A priori knowledge obtained from 
literature 
(e.g., \textit{IVIVE}) 
was used as reference values in parameter optimization to aid convergence to biologically plausible estimates.
While a global minimum in parameter optimization is not guaranteed, we find good biologically plausible estimates.
Our nested co-optimization approach allows us to partition parameters into independent and common parameters when using combined pre/post intervention data sets. 
This allows us to identify changes in physiological activities per person 
and thus differentiate the individuals based on the physiological activities that cannot be measured directly.

\mypar{Computational cost and time}
iPBPK-R 
was implemented using R leading to high computational cost (150,000 CPU hours total over the course of its development) and long run times (days per optimization run),  utilizing the \textit{Bridges} supercomputer at the Pittsburgh Supercomputing Center (PSC). 
Efficiency could be gained by switching to C++ at the cost of
substantial development effort.

\subsection{Comparison to Other Approaches}
\mypar{Classical prediction vs.\ estimation}
A  \textit{cocktail approach} where a subject takes multiple activity-specific probe drugs is the standard 
method to indirectly estimate different biological activities \cite{RN686} \cite{RN685}.
The group-wide activities associated with the respective drugs are statistically estimated,
in contrast to individual estimates in iPBPK-R.

Conventional \textit{population PBPK modeling}
is statistics-based modeling for an entire population.
It uses
a limitless number of parameters to capture all foreseen variabilities in various clinical scenarios and
IVIVE values are 
used as fixed input for simulations and prediction.
In population PBPK modeling researchers can assume some distributions on the property parameters 
and generate distributional parameter inputs using Monte Carlo approaches or similar methods. As a result the simulated drug concentration-time curves will vary broadly and have to be statistically summarized (median, 95\% CI, etc.). 

Due to the broader range of simulation curve outputs and the large number of input parameters, PBPK modeling is usually used to construct the statistical bounds of drug safety and/or efficacy. 
In contrast, iPBPK-R was developed for estimating multiple physiological parameters of individuals.
It enables multiple parameter estimation with a single probe via individual model fit.

\mypar{Mechanistic vs.\ statistical modeling}
iPBPK-R does not use distributional assumptions for 
parameters but uses them as reference values in the optimization procedure
to estimate individual parameters. 
Using a statistical population PBPK model would likely produce biased estimates since many co-dependent physiological parameters 
are treated as independent in their distributional assumptions.
IVIVE values 
are not necessarily clinically relevant population means, as they extrapolate \textit{in vitro} experimental values based on simple mathematical models.
In contrast we estimate
per-individual deviations from IVIVE values to overcome their limitation.

\mypar{Validation and model selection}
In population PBPK modeling, a model validation step is required to confirm a model's predictive ability when observed data become available. 
In contrast,
iPBPK-R uses a pre-determined model that is fit to individual observed data to estimate parameters given the model structure. 
This approach is inherently different from 
model 
validation for prediction: the model either works as is evident from good fit and biologically plausible parameter estimates, or a better PBPK model needs to be developed re-opening the PK modeling process.

With respect to model fitting, there is no formal statistical test to compare two sets of ODE parameters that have functional dependencies within each set. However, it is generally understood that such two sets of parameters for the same nonlinear system of ODEs result in quite different sets of solutions that will have obvious differences in model fit. 
The theory and result presented shows that our application examples were free from overfitting and estimation issues.
Our approach is not compatible with covariate selection in general PK modeling (e.g., nonlinear mixed models) where a covariate term is linealized and covariates are added or dropped based on some statistical threshold.

\mypar{PBPK software} 
A number of software tools are available to conduct PBPK simulations. 
The Simcyp Simulator by Certara USA, Inc. \cite{RN742} provides
PBPK modeling for a virtual populations of interest. 
NONMEM by ICON \cite{RN743} is FORTRAN-based biomathematical modeling software that allows users to explicitly specify mathematical models \cite{RN206}.
ADAPT 5 is another FORTRAN-based 
software \cite{RN744} \cite{RN745}.
These systems require programming in legacy languages like FORTRAN in combination with system-specific commands and features, which hampers more customized PBPK modeling and method research. 
Thus, iPBPK-R was developed using the free statistics-centered R system 
\cite{RN746}. 
Powerful optimization is available for model fitting, and parameter estimation across a wide range of observed data types (including breath rate data) can be implemented, as can 
parameter co-optimized across clinical intervention we use to correlate behavior before and after dialysis for kidney disease patients.

\section{Conclusions}
\label{sec:conclusions}
We presented iPBPK-R, a novel signal processing based \textit{indirect measurement} method, which is useful to simultaneously estimate multiple physiological parameters in individuals using clinical observed data. For this purpose, breath biopsy data is particularly well-suited as it can be modeled as the first derivative of drug concentration data and sampled at high clinical frequency to resolve early transients of drug behavior. 
The core of iPBPK-R is the joint estimation of multiple biological parameters via fitting of a nonlinear system of ODEs across all measurement data of an individual. We establish the mathematical foundations and computational framework of the parameter estimation and provide evidence that the parameter estimates obtained by iPBPK-R are stable and accurate.

We analyze the method's feasibility and establish its soundness, building on two published applications of iPBPK-R. Excellent model fits were achieved in individuals where multiple dependent parameters are estimated, and biological findings derived from iPBPK-R were compatible with independent biological experiments. Based on this early success with EBT we anticipate a path to eventual utilization of iPBPK-R in large-scale clinical trials, which may aid steps towards probe and biomarker development in personalized medicine. In future work other types of clinical datasets beyond EBT including multiple coupled drug concentration datasets per individual will be used.

\appendix
\section*{Appendix: Tables}

\begin{table}[h]
\centering
\abovedisplayskip=0pt
\belowdisplayskip=0pt
\begin{tabular} {m{14cm}}
\hline
Drug mass flows \\
\hline 
$q_{\overrightarrow{\IV:\VN}} =  M/t_0 \quad \mathrm{for} \ 0 \leq t \leq t_0,  \ \text{and}  \ 0 \ \mathrm{else}$
\\[-0ex]
$q_{\overrightarrow{\VN:\LG}} = \QTL\big(C_{\VN}-\frac{C_{\LG}}{\PTLG}\big)$,
$q_{\overrightarrow{\LG:\Art}} = \QTL\big(\frac{C_{\LG}}{\PTLG}- C_{\Art}\big)$,
$q_{\overrightarrow{\KD:\VN}} = \QKD\big(\frac{C_{\KD}}{\PTKD}-C_{\VN}\big)$
\\[-0ex]
$q_{\overrightarrow{\KD:\UR}} = \eGFR\frac{\fBL}{\PTKD}C_{\KD} +
             \mathrm{MM}_{\VmaxpgpKD,\KmpgpKD}\big(\frac{\fBL}{\PTKD}C_{\KD} \big)$,
$q_{\overrightarrow{\LC:\MB}} = \CLcyp\frac{\fLC}{\PTLCP}C_{\LC}$
\\[-0ex]
$q_{\overrightarrow{\Art:\KD}} = \QKD\big(C_{\Art}-\frac{C_{\KD}}{\PTKD}\big)$,
$q_{\overrightarrow{\OT:\VN}} = \QOT\big(\frac{C_{\OT}}{\PTOT}-C_{\VN}\big)$,
$q_{\overrightarrow{\Art:\OT}} = \QOT\big(C_{\Art}-\frac{C_{\OT}}{\PTOT}\big)$
\\[-0ex]
$q_{\overrightarrow{\OT:\DC}} = \mu_{\OT} C_{\OT}$,
$q_{\overrightarrow{\Art:\ES}} = \QLVAT C_{\Art}$,
$q_{\overrightarrow{\ES:\VN}} = \QLVVN C_{\VN|\ES}$
\\[-0ex]
$q_{\overrightarrow{\ES:\LC}} = \mathrm{MM}_{\VmaxupES,\KmupES}\big(\frac{\fES}{\PTESP}C_{\ES} \big)
+ \QESLC\big(\frac{\fES}{\PTESP}C_{\ES}
             -\frac{\fLC}{\PTLCP}C_{\LC}\big)$ \\[-0ex]
$q_{\overrightarrow{\LC:\BE}} = \mathrm{MM}_{\VmaxpgpLC,\KmpgpLC}\big(\frac{\fLC}{\PTLCP}C_{\LC} \big)
+ \mathrm{MM}_{\VmaxmrpLC,\KmmrpLC}\big(\frac{\fLC}{\PTLCP}C_{\LC} \big)$
 \\
\hline
\end{tabular}
\caption{Mass flows of $^{14}$C-erythromycin in Fig.~\ref{fig:ebtmb1}.
\label{table:3}}
\end{table}

\begin{table}[h]
\centering
\abovedisplayskip=0pt
\belowdisplayskip=0pt
\begin{tabular}{p{3cm} p{11cm} }
\hline
Parameter & definition \\
\hline 
$\QTL$, $\QESLC$  & total blood flow and  passive diffusion Comp. $\ES$ and $\LC$
\\[0ex]
$\QLVVN$, $\QLVAT$ & blood flow Compartment $\ES$ to $\VN$ and Compartment $\Art$ to $\ES$\\[0ex]
$\QOT$, $\QKD$ & blood flow in and out of Compartment $\OT$ and $\KD$ \\[0ex]
$\mathrm{P}_i$, 
$\mathrm{P}_{i\mathrm{p}}$ & partition coefficient of Compartment $i$ to blood and plasma
\\[0ex]
$\mathrm{f}_i$, $\mathrm{f}_{i\mathrm{b}}$  & fraction unbound in Compartment $i$ and blood\\[0ex]
$\eGFR$ & estimated GFR, scaled with GFR filtration fraction $\delta$  \\[0ex]
$\VmaxupES, \, \VmaxpgpLC, \, \VmaxmrpLC$ & maximum velocity of drug transporter a, b, and c 
\\[0ex]
$\KmupES, \, \KmpgpLC, \, \KmmrpLC$ & Michaelis-Menten constant of drug transporter a, b, and c 
\\[0ex]
$\eta$ & distribution ratio of transporter b in Compartment $\KD$ to $\LC$\\[0ex]
$\mu$ & exponential decay parameter of the parent drug \\[0ex]
$\CLcyp$ &  CYP3A4 clearance (calculated IVIVE value) \\[0ex]
$k_1$ & distribution rate of $\mathrm{H^{14}CO}^-_3$ from liver to pool (0.0~\cite{RN622}) \\[0ex]
$k_2$ & distribution rate of $\mathrm{H^{14}CO}^-_3$ from  pool to liver (0.0~\cite{RN622}) \\[0ex]
$k_3$ & excretion rate constant of $\mathrm{^{14}CO}_2$ in the liver cell\\[0ex]
$\phi$ & conversion of  $\mathrm{^{14}CO}_2$ from mol to Ci (here, $\phi = 1$) \\
\hline
\end{tabular}
\caption{Drug flows parameters of $^{14}$C-erythromycin.}\label{table:2}
\end{table}

\begin{table}[h]
\centering
\abovedisplayskip=0pt
\belowdisplayskip=0pt
\begin{tabular}{p{3cm} p{11cm} }
\hline
Adj. factor & Definition: \textit{adjustment factor of} \\
\hline 
$\alpha_{\OT}$ & exponential decay in Other Organs compartment \\[0ex]
$\alpha_{\LC}$ & CYP3A4 activity in Liver Cell compartment \\[0ex]
$\beta_{\ES}$ & drug transporter a in the liver \\[0ex]
$\beta_{\LC,1}$ & drug transporter b in the liver \\[0ex]
$\beta_{\LC,2}$ & drug transporter c in the liver \\[0ex]
$\gamma_{\OT}$ & partition coefficient of Other Organs to blood \\[0ex]
$\zeta_{\LC}$ & volume of Liver Cell compartment \\[0ex]
$\kappa_{\OT\Art}$ & arterial blood flow into Other Organs compartment \\[0ex]
$\kappa_{\ES\LC}$ & passive diffusion between ES and Liver Cell comp. \\
\hline
\end{tabular}
\caption{Adjustment factors in iPBPK-R for EBT.}\label{table:4}
\end{table}

\begin{table}[h!]
\centering
\abovedisplayskip=0pt
\belowdisplayskip=0pt
\begin{tabular}{ll}
\hline
Entries of $\mathbf{X}$, $\mathbf{\tilde{Y}}$ and $\mathbf{D}$\\
\hline 
$x_{1,1} = -2\frac{\QTL}{\VLG\PTLG}$, $x_{1,2}=\frac{\QTL}{\VLG}$, $x_{1,5}=\frac{\QTL}{\VLG}$,
$x_{2,1} = \frac{\QTL}{\VVN\PTLG}$,
 \\[1ex]
$x_{2,2}=-\frac{\QTL+\QOT+\QKD}{\VVN}$, $x_{2,3}=\frac{\QOT}{\gamma_{\OT}\VVN\PTOT}$,
 \\[1ex]
$x_{2,4} = \frac{\QKD}{\VKD\PTKD}$, $x_{2,6}=\frac{\QLVVN}{\VVN\PTES}$
\\[1ex]
$x_{3,2} = \frac{\QOT}{\VOT}$, $x_{3,3}=-\frac{(\kappa_{\OT\Art}+1)\frac{\QOT}{\gamma_{\OT}\PTOT}+\alpha_{\OT}\mu_{\OT}}{\VOT}$,
 \\[1ex]
$x_{3,5}=\frac{\kappa_{\OT\Art}\QOT}{\VOT}$,
\\[1ex]
$x_{4,2} = \frac{\QKD}{\VKD}$, $x_{4,4}=-\frac{2\QKD+\eGFR\fBL}{\VKD\PTKD}$, $x_{4,5}=\frac{\QKD}{\VKD}$,
\\[1ex]
$x_{5,1} = \frac{\QTL}{\VAT\PTLG}$,
$x_{5,3}=\frac{\kappa_{\OT\Art}\QOT}{\gamma_{\OT}\VAT\PTOT}$, $x_{5,4}=\frac{\QKD}{\VAT\PTKD}$,
 \\[1ex]
$x_{5,5} = -\frac{\QTL+\kappa_{\OT\Art}\QOT+\QKD}{\VAT}$, $x_{5,6}=- \frac{\QLVAT}{\VAT}$,
\\[1ex]
$x_{6,5} = \frac{\QLVAT}{\VESef}$, $x_{6,6}=- \frac{1}{\VESef}\left(\frac{\kappa_{\ES\LC}\QESLC\fES}{\PTESP} + \frac{\QLVVN}{\PTES}\right)$, $x_{6,7}=\frac{\kappa_{\ES\LC}\QESLC\fES}{\VESef\PTLCP}$,
\\[1ex]
$x_{7,6}=\frac{\kappa_{\ES\LC}\QESLC\fES}{\zeta_{\LC}\VLC\PTESP}$, $x_{7,7}=- \frac{(\kappa_{\ES\LC}\QESLC + \alpha_{\LC}\CLcyp)\fLC}{\zeta_{\LC}\VLC\PTLCP}$,
$x_{9,3}=\frac{\alpha_{\OT}\mu_{\OT}}{\VOT}$,
 &
 \\[1ex]
$x_{10,4}=\frac{\eGFR\fBL}{\VKD\PTKD}$, $x_{11,7}=\frac{\alpha_{\LC}\CLcyp\fLC}{\zeta_{\LC}\VLC\PTLCP}$,
\\[1ex]
$x_{13,7}=\frac{\alpha_{\LC}\CLcyp\fLC}{\zeta_{\LC}\VLC\PTLCP}$,
$x_{13,13}=-k_3$,
$x_{14,13}=\phi k_3$
 \\[1ex]
$y_{4,4} = -\mathrm{MM}_{\beta_{\LC,1}\VmaxpgpKD,\KmpgpKD}$ \\[1ex]
$y_{6,6} = -\mathrm{MM}_{\beta_{\ES}\VmaxupES,\KmupES}$ \\[1ex]
$y_{7,6} = \mathrm{MM}_{\beta_{\ES}\VmaxupES,\KmupES}$ \\[1ex]
$y_{7,7} = -(\mathrm{MM}_{\beta_{\LC,1}\VmaxpgpLC,\KmpgpLC} + \mathrm{MM}_{\beta_{\LC,2}\VmaxmrpLC,\KmmrpLC})$ \\[1ex]
$y_{10,4} = \mathrm{MM}_{\beta_{\LC,1}\VmaxpgpKD,\KmpgpKD}$ \\[1ex]
$y_{12,7} = \mathrm{MM}_{\beta_{\LC,1}\VmaxpgpLC,\KmpgpLC} + \mathrm{MM}_{\beta_{\LC,2}\VmaxmrpLC,\KmmrpLC}$  \\[1ex]
$d_{4} = \frac{\fBL}{\VKD\PTKD} $, $d_6 = \frac{\fES}{\VESef\PTESP}$, $d_7 = \frac{\fLC}{\zeta_{\LC}\VLC\PTLCP}$\\[1ex]
\hline
\end{tabular}
\caption{Entries of matrices $\mathbf{X}$, $\mathbf{\tilde{Y}}$, and $\mathbf{D}$.}\label{table:8}
\end{table}

\clearpage

\bibliographystyle{plain}
\bibliography{references_iPBPKR}

\end{document}